\documentclass[%
 preprint,
 amsmath,amssymb,
 aps, physrev,
]{revtex4-2}

\usepackage{graphicx}
\usepackage{dcolumn}
\usepackage{bm}

\begin{document}


\title{\textbf{Response of Confined Mesoscale Flames to Acoustic Forcing} 
}%

\author{Akhil Aravind}
\author{Balasundaram Mohan}
\author{Saptarshi Basu}%
 \email{Contact author: sbasu@iisc.ac.in}
\affiliation{%
 Department of Mechanical Engineering, Indian Institute of Science, Bengaluru, India.}%


\begin{abstract}
Mesoscale combustion plays a vital role in applications ranging from portable power generation to micro-propulsion systems. This study investigates the response of mesoscale flames to upstream acoustic forcing across three characteristic flame regimes: Flames with Repetitive Extinction and Ignition (FREI), Stable Flames (SF), and Propagating Flames (PF). Acoustic forcing is applied over a broad range of pressure intensities at the first harmonic of the combustor tube. At low forcing intensities, all the flame regimes exhibit oscillations at the forcing frequency without qualitative changes in their respective flame behaviours. However, as the acoustic pressure increases, all flame regimes transition to a common unsteady regime characterised by periodic ignition-extinction cycles (resembling FREI) alongside strong oscillations at the forcing frequency. This response is identified as the oscillating FREI (O-FREI) regime, in which increasing forcing intensity leads to a reduction in the mean heat release rate, an amplification of oscillations at the forcing frequency, and a drop in the Strouhal number associated with the ignition-extinction cycle. The response characteristics across all flame regimes are examined in detail and are further complemented using reduced-order mathematical models.
\end{abstract}

\maketitle


\section{Introduction} \label{sec:intro}

Micro power generation systems are increasingly being explored as potential alternatives to batteries for powering micro-electro-mechanical systems (MEMS), with applications ranging from biomedical devices to aerial and space vehicles. The appeal of these systems stems from the substantially higher gravimetric energy density of hydrocarbon fuels compared to conventional electrochemical storage, combined with shorter recharge times \cite{fernandez-pello_micropower_2002, ju_microscale_2011}. Additionally, micro- and mesoscale combustion has gained significant relevance in the field of quantitative chemistry following the development of micro-reactors \cite{ZIMMERMANN2002285}.

In mesoscale combustors, the characteristic length scale of the combustor is comparable to the flame thickness, while in microscale combustors, it is even smaller \cite{ju_microscale_2011}. Reducing the combustor dimensions to such length scales introduces several challenges, primarily due to enhanced wall heat losses that can lead to thermal quenching of the flame, as well as intensified flame-wall interactions that promote radical quenching and eventual flame extinction. To mitigate these effects in practical combustors operating at these length scales, the combustor walls are actively heated \cite{lloyd_burner_1974}, either through recirculation of hot combustion products or via an external heat source. The coupled interplay of convective transport, wall heat losses, flame-wall interactions, and duct acoustics gives rise to a range of unique flame behaviours in mesoscale combustors that are not typically observed in conventional, large-scale combustion systems. A brief overview of the commonly reported flame regimes in such systems is presented below.

One of the early fundamental studies on microscale combustion was carried out by Maruta \textit{et al.} \cite{maruta_characteristics_2005, maruta_micro_2011}, who reported a range of flame behaviours in premixed methane–air flames stabilised in a 2 mm quartz tube. These included stable flames, flames exhibiting repetitive extinction and ignition (FREI), pulsating flames, and mixed regimes combining features of FREI and pulsation.
The FREI regime is of particular interest because the flame does not stabilise at a fixed location. Instead, it undergoes limit-cycle oscillations involving ignition, upstream propagation, extinction, and re-ignition. This behaviour arises from the competition between upstream convection, local wall heat losses, and local heat release, with the wall temperature gradient playing a central role in setting the spatial extent of the oscillation.
Subsequent studies reported similar flame behaviours across broad operating conditions, including variations in reactant composition, combustor wall properties, wall spacing, and imposed wall temperature profiles \cite{fan_experimental_2009-1, di_stazio_combustion_2016, pizza_dynamics_2008, kang_versatile_2021}.

Complementary theoretical studies have provided further insight into these behaviours. \cite{maruta_characteristics_2004} used linear stability analysis to identify the transition between stable flames and unsteady FREI regimes. Building on this, \cite{minaev_nonlinear_2007} reduced the one-dimensional convection–diffusion formulation by assuming constant density, constant mixture properties, and an infinitesimally thin reaction zone. This led to analytical expressions for flame temperature and position in the stable flame regime, along with a reduced-order dynamical system that reproduces the qualitative features of FREI. This framework reduces the full transport problem to a low-dimensional representation while retaining the key physical mechanisms and remaining analytically tractable.

The observed flame regimes are strongly dependent on combustor configuration. \cite{aravind_dynamics_2025} recently identified an additional periodically repeating regime — propagating flames — in lean methane–air mixtures in an acoustically closed–open mesoscale combustor. In this regime, the flame traverses the full combustor length while exhibiting thermoacoustic coupling during the propagation phase, demonstrating that acoustic waves can significantly modify flame behaviour even in the absence of deliberate external forcing. Similar thermoacoustic coupling of propagating premixed flames in quiescent mixtures confined to slender, mesoscale channels has also been reported by \cite{Flores-Montoya_Muntean_Sánchez-Sanz_Martínez-Ruiz_2022}. 

Geometric modifications of the combustor flow domain have also been explored as a means of influencing mesoscale flame dynamics, including the use of bluff bodies \cite{wan_experimental_2012, sheykhbaglou_numerical_2025} and wavy channels \cite{resende_flame_2023, gao_numerical_2023}. Such geometric alterations can induce vortex-shedding structures that interact with the flame and potentially destabilise it, particularly when the shedding frequency approaches the natural acoustic harmonic of the combustor.

Building upon these considerations, it is essential to understand the response of mesoscale flames when excited at the natural harmonic of the combustor—a situation that can arise in practice due to flow instabilities or coupling with the combustor acoustics.
While analogous forced-response studies have been extensively conducted in large-scale combustion systems \cite{ducruix2000theoretical, noiray2008unified}, where flame responses in both linear and nonlinear forcing regimes have been systematically characterised through flame transfer functions and describing functions, such investigations remain largely unexplored at micro and mesoscales. Importantly, insights from large-scale systems cannot be directly extrapolated to micro/mesoscale combustors, as the underlying flame regimes — particularly FREI and propagating flames — have no direct analogues in conventional combustors, and the dominant physical mechanisms differ fundamentally at reduced length scales. Addressing this gap, the present work constitutes a first-of-its-kind study examining the response of mesoscale flames subjected to external forcing at the natural harmonic of the combustor, spanning all three characteristic flame regimes — FREI, stable flames, and propagating flames — over a wide range of forcing amplitudes.

\section{Experimental Facility} \label{sec:Expt}

This section details the experimental facility and the data processing techniques used in the study.

\subsection{Experimental setup}

\begin{figure}
    \centering
    \includegraphics[width=0.75\textwidth]{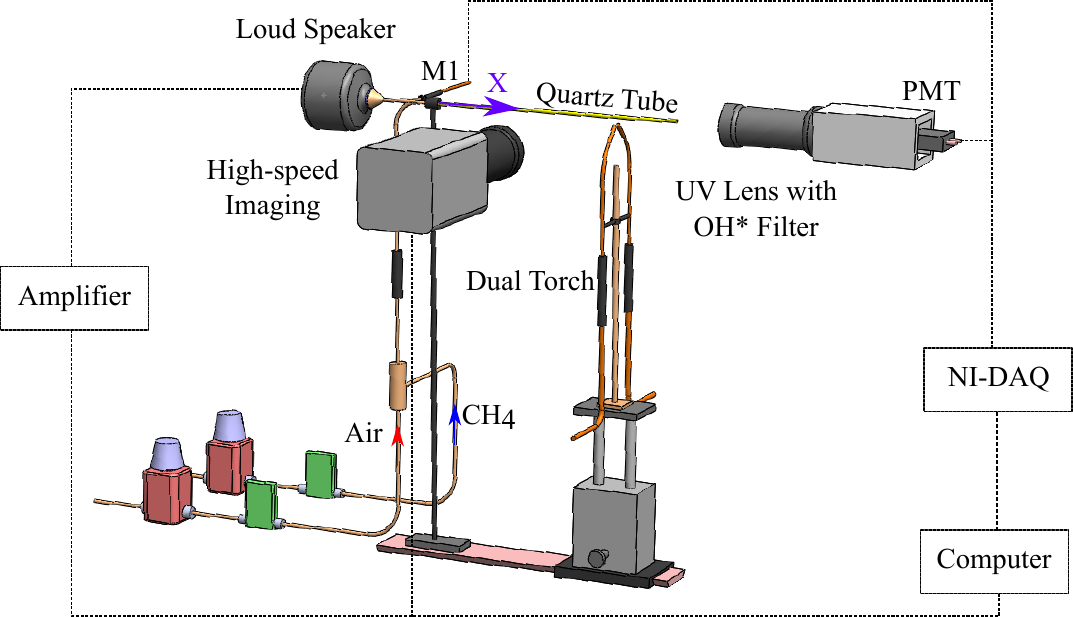}
    \caption{Schematic of the experimental facility. In the figure, the quartz tube is oriented along the X-axis, with $X=0$ corresponding to the upstream end of the tube.}
    \label{fig:expt}
\end{figure}

The experimental setup consists of a long quartz tube (length, $L=320$ mm) with an inner diameter of 5 mm and a wall thickness of 1 mm, which functions as an optically accessible mesoscale combustor. The upstream end of the tube is connected to a constricted four-way junction, whereas the downstream end is open to the atmosphere. At the upstream junction, the quartz channel is axially coupled to a monotone sound driver. One lateral port of the junction is equipped with a microphone (M1), and the other lateral port delivers the premixed reactant mixture into the quartz channel, as shown in Fig. \ref{fig:expt}. The flow rates of methane (fuel) and air (oxidizer) are controlled using two Bronkhorst Flexi-Flow Compact mass flow controllers. The reactants are mixed in a premixing chamber before introducing them at the upstream end of the tube. In the figure, $X = 0$ corresponds to the upstream end of the tube, while $X = L$ corresponds to the downstream end, which is open to the atmosphere.

\begin{figure}
    \centering
    \includegraphics[width=0.85\textwidth]{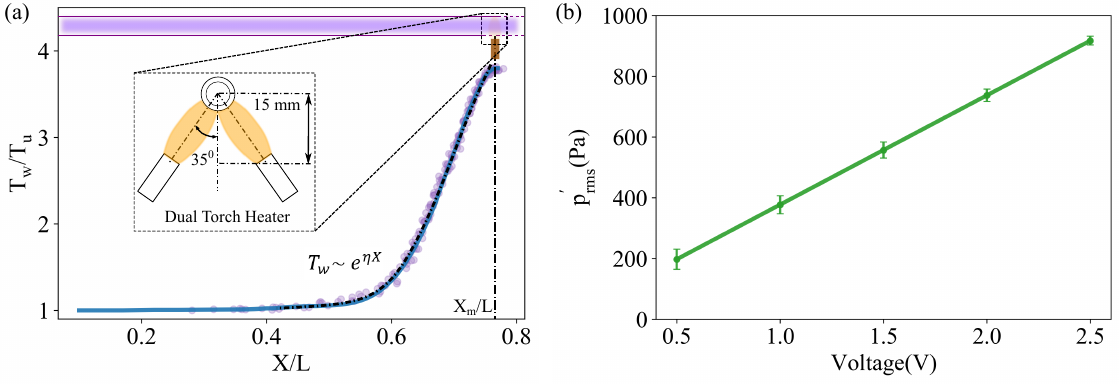}
    \caption{(a) Inner-wall temperature profile ($T_w$), normalised by the unburnt gas temperature ($T_u$), measured along the lower periphery of the quartz tube. Here, $L$ denotes the length of the combustor tube. (b) Root-mean-square (rms) magnitude of the imposed pressure fluctuations measured upstream of the combustor tube using microphone M1, for different amplifier control voltages.}
    \label{fig:expt_profile}
\end{figure}

During operation, the quartz tube is externally heated using a dual-torch system positioned 80 mm from the downstream end ($X=X_m$ in Fig. \ref{fig:expt_profile}a), as shown in Fig. \ref{fig:expt}. The heating system comprises two non-premixed liquefied petroleum gas (LPG) jet flames oriented such that the quartz tube lies at the stagnation plane formed by the opposing jets (Fig. \ref{fig:expt_profile}a). The resulting inner wall temperature ($T_w$, normalised by the unburnt gas temperature $T_u$), measured along the lower periphery of the tube using a 2 mm K-type thermocouple, is shown in Fig. \ref{fig:expt_profile}(a). 
Owing to the closer proximity of the dual-torch arrangement to the lower periphery of the quartz tube (relative to the upper periphery), the resulting wall temperature distribution exhibits radial non-uniformity. Our previous study \cite{aravind_dynamics_2025} employing a similar heating configuration has shown that the maximum fractional temperature difference between the upper and lower peripheries occurs in the vicinity of the heater and remains within approximately 2 \% of the local measured temperature. This asymmetry causes the flame to auto-ignite preferentially at the lower periphery of the tube, at axial locations close to the external burner. However, as the flame propagates upstream, away from this region, it rapidly attains a nearly uniform structure across the tube cross-section.

External forcing was applied using a monotone sound driver (Ahuja AU-50XT) powered by an Ahuja SSA-100M amplifier. Sinusoidal signals generated via a Python-based script were fed into the amplifier, inducing periodic upstream flow fluctuations. The study investigates the response dynamics of three distinct mesoscale flame regimes: stable flames (SF), flames exhibiting repetitive extinction and ignition (FREI), and propagating flames (PF). 
Experiments were conducted at three operating conditions (Table 1), selected to represent each flame regime.
\begin{table}
\vspace{-1\baselineskip}
\centering
\caption{Operating conditions of the flame regimes in terms of the mean flow velocity ($\overline{u}$), Reynolds number ($Re$) and equivalence ratio ($\Phi$)}
\label{tab:flame_regimes}
\begin{tabular}{lccc}
\hline
Flame Regime & $\overline{u}$ (m/s) & $Re$ & $\Phi$ \\
\hline
FREI & 0.3 & 96  & 1.0 \\
SF & 0.7 & 224 & 0.8 \\
PF & 0.3 & 96  & 0.8 \\
\hline
\end{tabular}
\end{table}

During the experiments, the amplitude of the imposed upstream forcing was systematically varied, while the forcing frequency was fixed at the first harmonic of the combustor tube. Forcing amplitude was adjusted by varying the amplifier control voltage from 0 to 2.5 V. The corresponding RMS pressure fluctuations measured by microphone M1 is presented in Fig. \ref{fig:expt_profile}(b).

The flame response was characterised using high-speed OH* chemiluminescence measurements acquired with a Hamamatsu H11526-110-NF photomultiplier tube (PMT) fitted with an OH* bandpass filter ($\sim 310 \pm 10$ nm) and aligned along the channel axis (Fig. \ref{fig:expt}). Simultaneous high-speed pressure measurements were obtained using a PCB 377B26 microphone (M1) located at the upstream constricted end of the combustor. PMT and microphone signals were recorded at a sampling rate of 16,384 Hz using a data acquisition system (NI DAQ PCI-6251).
Simultaneous flame imaging was performed using a Photron SA5 high-speed camera at 500 frames per second, with a spatial resolution of 200 $\mu$m per pixel, an exposure time of 2 ms, and an image size of $1024 \times 96$ pixels. The frame rate was limited by the low flame luminosity and the corresponding requirement for long exposure times. 
It should be noted that the imaging data were used exclusively to determine the mean flame position; all subsequent analysis is based primarily on the OH* chemiluminescence and pressure measurements.

\subsection{Data Processing}

The time-series data acquired from the pressure sensors and the photomultiplier tube were processed in \textsc{MATLAB}. The signals were filtered using the Savitzky--Golay filter to reduce noise while preserving the salient features of the signal. Fast Fourier Transform (FFT) analysis was then performed to identify the dominant time scales present in the signals and their corresponding amplitudes. In the sections that follow, the variation of these dominant time scales and their amplitudes with the imposed upstream forcing pressure is discussed in detail.

The flame imaging data acquired using the high-speed camera were processed using the open-source software \textsc{ImageJ}. The flame boundary was identified using Otsu’s thresholding technique, enabling binarisation of the images and separation of the flame from the background. Once the flame boundary was obtained, the centroid of the identified flame region was estimated and tracked as the flame position. 
This technique was employed in our previous studies to track the dynamics of flame front evolution when subjected to impulsive flow fields \cite{aravind_response_2025}. 
The temporal evolution of the flame position was subsequently analysed in conjunction with the OH* chemiluminescence and pressure signals to characterise the flame dynamics. 

The quantitative measurements (e.g., mean heat release rate, heat release rate fluctuation amplitude, and ignition-extinction Strouhal number) of flame dynamics presented in the following sections correspond to results averaged over at least three independent repetitive trials. The error bars represent the standard deviation across these trials.

\section{Global Observations} \label{sec:glo_obs}
This section describes the qualitative response characteristics of the different flame regimes to acoustic forcing at the natural harmonic of the tube ($f_n$) over the range of forcing amplitudes presented in Fig. \ref{fig:expt_profile}(b). 

\begin{figure}
    \centering
    \includegraphics[width=0.75\textwidth]{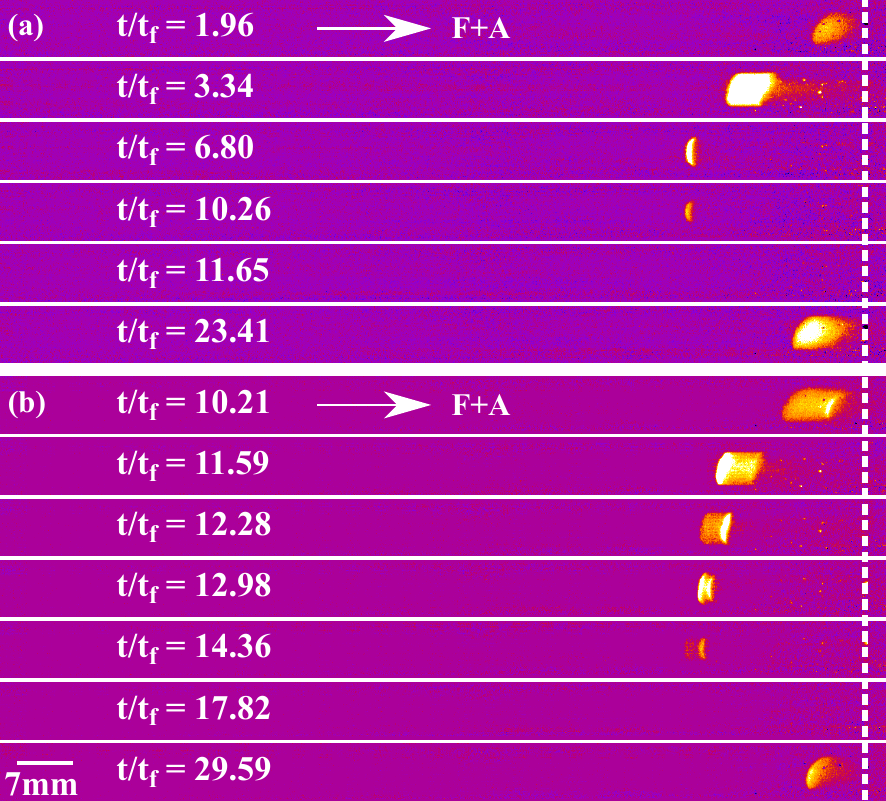}
    \caption{(a) Image sequence illustrating the dynamics of flames with repetitive extinction and ignition (FREI) at $Re=96$ and $\Phi=1.0$. (b) Dynamics of FREI under imposed upstream pressure fluctuations with an rms amplitude of 197 Pa (oscillating FREI). In the figure, $t_{f}$ denotes the timescale associated with the forcing frequency. The dotted vertical line indicates the position of the external heater. Video illustrations in Supplementary Movies M1 and M2.}
    \label{fig:FREI_Obs}
\end{figure}

The premixed reactants enter the combustor tube at the upstream end and flow toward the heated downstream section. During this transit, the mixture absorbs heat from the combustor walls, leading to an increase in its mean temperature. The mean temperature of the premixed reactants closely follows the spatial profile of the imposed wall temperature, with a slight downstream shift that increases with increasing mean flow velocity, as shown in our previous study \cite{aravind_dynamics_2025}. 
The plots illustrating the spatial profile of the mean flow temperature of the reactant mixture is presented in Appendix A.
When the mixture temperature exceeds the auto-ignition threshold, ignition occurs, typically at axial locations close to the external heater. The subsequent flame behaviour depends strongly on the operating flame regime.

\begin{figure}
    \centering
    \includegraphics[width=0.75\textwidth]{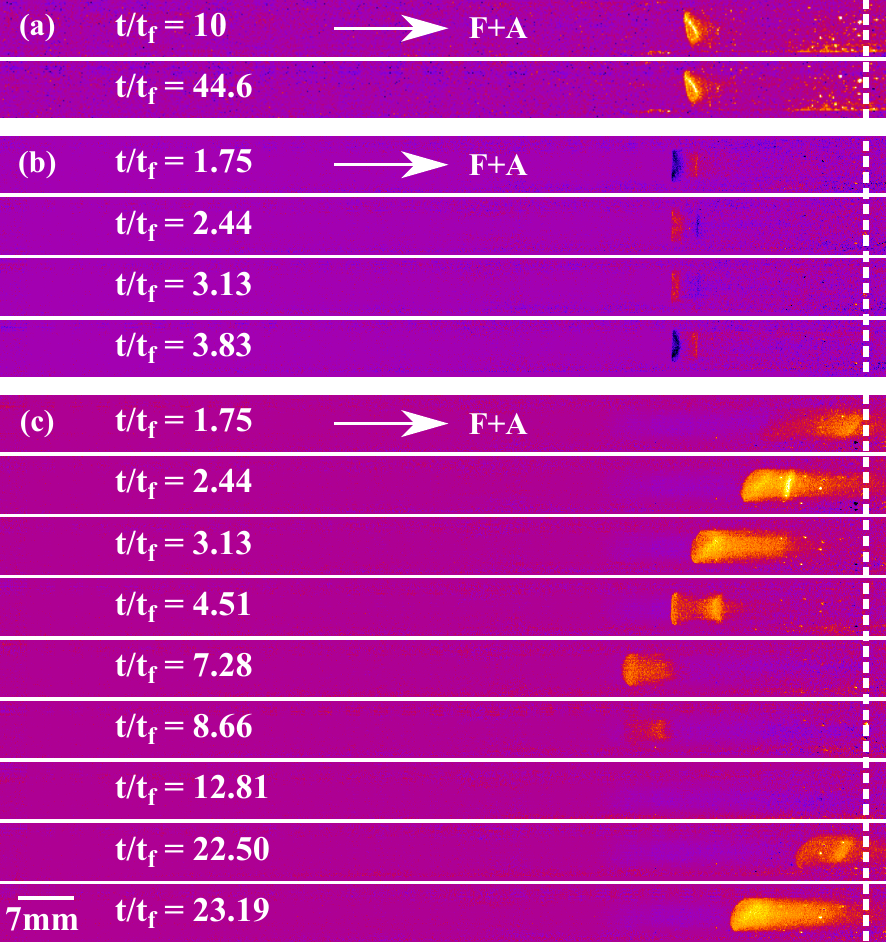}
    \caption{(a) Image sequence illustrating the dynamics of a stable flame (SF) at $Re=224$ and $\Phi=0.8$. (b) Dynamics of the stable flame under imposed upstream pressure fluctuations with an rms amplitude of 197 Pa (oscillating SF). (c) 377 Pa (oscillating FREI). Video illustrations in Supplementary Movies M3, M4 and M5.}
    \label{fig:SF_Obs}
\end{figure}

\begin{figure}
    \centering
    \includegraphics[width=0.75\textwidth]{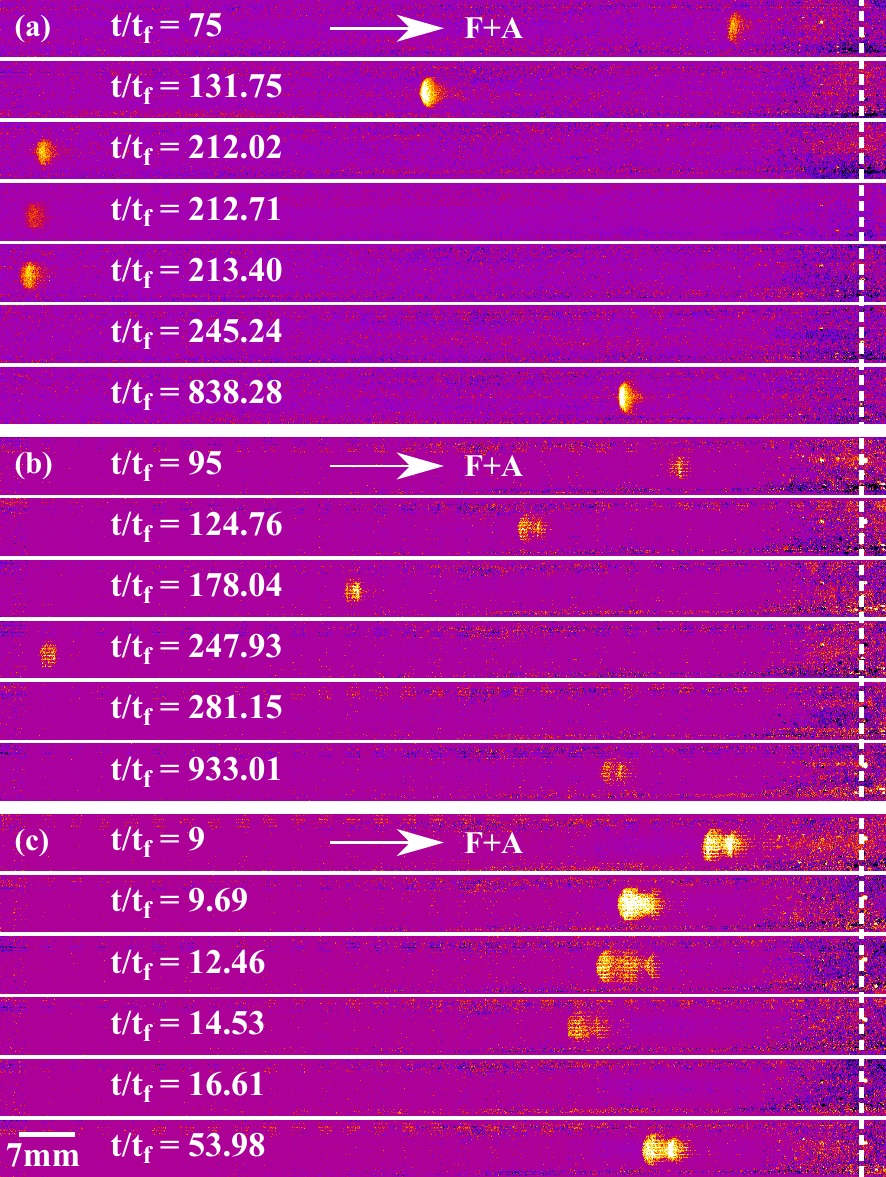}
    \caption{(a) Image sequence illustrating the dynamics of a propagating flame (PF) at $Re=96$ and $\Phi = 0.8$. (b) Dynamics of the propagating flame under imposed upstream pressure fluctuations with an rms amplitude of 197 Pa (oscillating PF). (c) 377 Pa (oscillating FREI). Video illustrations in Supplementary Movies M6, M7 and M8.}
    \label{fig:PF_Obs}
\end{figure}

In flames exhibiting repetitive extinction and ignition (FREI), the auto-ignited flame propagates upstream, extinguishes after a characteristic travel distance, and subsequently re-ignites after a fixed time interval, resulting in a periodic ignition–extinction cycle (Fig. \ref{fig:FREI_Obs}a). In contrast, in the stable flame (SF) regime, the auto-ignited flame propagates upstream and stabilizes at a fixed axial location determined by the wall temperature profile (Fig. \ref{fig:SF_Obs}a).
Propagating flames (PF), observed under fuel-lean conditions, continue to travel toward the upstream end of the tube following auto-ignition, while exhibiting thermoacoustic oscillations during the propagation phase (Fig. \ref{fig:PF_Obs}a). The flame is extinguished only at the constricted upstream end of the combustor ($X=0$). Similar to the FREI regime, PF dynamics are periodic, with the ignition–extinction cycle repeating over a well-defined time period.
The characteristics of the FREI and PF regimes across the parametric space of $Re$ and $\Phi$ have been documented in detail in our previous study \cite{aravind_dynamics_2025}.

Our previous work \cite{aravind_dynamics_2025} demonstrated that the thermoacoustic coupling frequency in the propagating flame (PF) regime is close to the first harmonic of the tube and agrees well with estimates obtained from frequency-domain analysis of the combustor tube, modelled as an open–closed system with appropriate mixture temperatures. Accordingly, the first natural harmonic of the tube ($f_n$) is inferred from the experimentally observed thermoacoustic coupling frequency in the PF regime and is approximately 346 Hz for the present configuration.

When flames with repetitive extinction and ignition (FREI) are forced at the frequency of $f_n$, they exhibit oscillations at the forcing frequency while retaining their cyclic sequence of ignition, upstream propagation, extinction, and re-ignition (Fig. \ref{fig:FREI_Obs}b). The resulting dynamics resemble a superposition of FREI behaviour with oscillations about the local mean at the forcing frequency (evident from the heat release rate signals presented in Section \ref{subsec:FR}). This is reflected in the flame images presented in Fig. \ref{fig:FREI_Obs}(b). This response regime is hereafter referred to as the oscillating FREI (O-FREI) regime. The flame maintains this qualitative behaviour over the range of forcing amplitudes explored in the present study. 
Quantitative variations in the dominant timescales and their associated amplitudes in the O-FREI regime, as a function of the imposed forcing pressure, are discussed in section \ref{subsec:FR}. 

When the stable flame is forced at the first harmonic of the tube at an upstream forcing pressure of 197 Pa (rms), it oscillates about its mean axial position, accompanied by corresponding oscillations in the heat release rate at the forcing frequency. This response is hereafter referred to as the oscillating stable flame (O-SF) regime and is illustrated in Fig. \ref{fig:SF_Obs}(b).
With a further increase in forcing pressure, the flame transitions to the previously described oscillating FREI (O-FREI) regime, characterised by periodic cycles of ignition, upstream propagation, extinction, and re-ignition (Fig. \ref{fig:SF_Obs}c), superimposed with local oscillations at the forcing frequency. Beyond this transition, the flame retains O-FREI behaviour over the entire range of higher forcing pressures investigated.

When the propagating flame is forced at the first harmonic frequency ($f_n$), it retains its global behaviour—complete upstream propagation followed by re-ignition after a characteristic time interval—at a forcing pressure of 197 Pa (rms). In this regime, the flame exhibits an increased amplitude of thermoacoustic oscillations during the propagation phase. This response is hereafter referred to as the oscillating propagating flame (O-PF) regime, to distinguish it from its unforced behaviour, and is illustrated in Fig. \ref{fig:PF_Obs}(b).
With further increase in forcing amplitude, the flame transitions to the oscillating FREI (O-FREI) regime (Fig. \ref{fig:PF_Obs}c) and retains this behaviour over the entire range of higher forcing pressures explored in the present study.

It is interesting to note that all the flame regimes, when forced beyond a critical forcing threshold, transition to the oscillating FREI (O-FREI) regime. The characteristics of this regime are explored in detail in the section \ref{subsec:FR}.

\section{Estimation of the forcing velocity fluctuation scale} \label{sec:vel_est}
\begin{figure}
    \centering
    \includegraphics[width=0.75\textwidth]{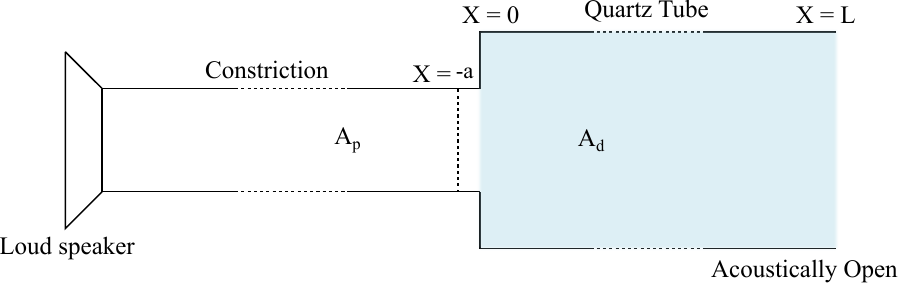}
    \caption{Schematic illustrating the area change between the constricted section, where the pressure is measured ($X=-a$), and the quartz tube, where the flame is stabilised.}
    \label{fig:acoustic_schematic}
\end{figure}

This section develops an estimate of the acoustic velocity fluctuations imposed on the flame by the external forcing field. The microphone M1 is located at the four-way upstream junction and is coupled to the quartz tube (cross-sectional area $A_d$) via a 2 mm constriction (cross-sectional area $A_p$), as illustrated schematically in Fig. \ref{fig:acoustic_schematic}. Because M1 measures pressure fluctuations on the constriction side of this junction ($X=-a$; Fig. \ref{fig:acoustic_schematic}), the measured signal must be corrected for the acoustic effect of the area change before it can be related to the acoustic velocity field inside the tube.

The area ratio $(A_p/A_d)$ is approximately 0.16, and therefore the junction constitutes an area discontinuity with a strong impedance mismatch. Assuming linear one-dimensional plane-wave propagation \cite{munjal1987acoustics}, the acoustic pressure and velocity fields in each duct may be written as superpositions of forward- and backward-travelling waves:
\begin{align*}
    p_i = & a_i e^{j(\omega t - kX)} + b_i e^{j(\omega t + kX)} \\
    u_i = & \frac{1}{\rho c} \left( a_i e^{j(\omega t - kX)} - b_i e^{j(\omega t + kX)} \right), \qquad i=1,2
    \tag{1}
\end{align*}
where $a_i$ and $b_i$ denote the amplitudes of the forward- and backward-travelling waves, respectively, and subscripts 1 and 2 refer to the constriction and the quartz tube (Fig. \ref{fig:acoustic_schematic}). Here, $j=\sqrt{-1}$ denotes the imaginary unit. Determining the acoustic field requires solving for four unknowns: $a_1$, $b_1$, $a_2$, and $b_2$, and therefore requires four additional equations relating these variables. Since the forcing is applied at the frequency of 346 Hz ($f_n$), $\omega$ and $k$ are known, where $\omega = 2\pi f_n$ and $k=\omega/c$, with $c$ denoting the speed of sound in the medium.

Two physical conditions must hold simultaneously at the junction ($X=0$): acoustic pressure must be continuous, and acoustic volume flow rate must be conserved \cite{munjal1987acoustics}. These yield,
\begin{gather*}
    a_1 + b_1 = a_2 + b_2 \\
     \frac{A_p}{\rho c} (a_1 - b_1) = \frac{A_d}{\rho c} (a_2 - b_2)
    \tag{2}
\end{gather*}
where $\rho$ is the density of the medium. Since the temperature of the medium remains approximately unchanged across the area discontinuity, the quantity $\rho c$ is assumed to be identical in the constriction and the quartz tube.

The far end of the tube ($X=L$) is open to the atmosphere, implying that the acoustic pressure at this location is zero. This yields,
\begin{gather*}
    a_2 e^{-jkL} + b_2 e^{jkL} = 0
    \tag{3}
\end{gather*}
The final condition is obtained from the microphone measurement. The microphone records the acoustic pressure at $X=-a$, yielding
\begin{gather*}
    (a_1 e^{jka} + b_1 e^{-jka})e^{j \omega t} = p_{mic} = p_a e^{j \omega t}
    \tag{4}
\end{gather*}
Solving for the unknown variables in Eq. (1) using the conditions specified in Eqs. (2)-(4), and subsequently evaluating the acoustic velocity field inside the quartz tube, yields
\begin{gather*}
    u_2 (x,t) = Re\left( \frac{p_a}{\rho c} \frac{ \left[ e^{jk(L-X)} + e^{-jk(L-X)} \right]}{\left[ j \sin(kL)\cos(ka) + j \frac{A_d}{A_p} \cos(kL)\sin(ka) \right]}  e^{j \omega t} \right)
    \tag{5}
\end{gather*}

It is to be noted that the above estimation neglects viscous and thermal losses across the constriction. Therefore, the calculated values represent an upper bound on the velocity fluctuation amplitude. 
Due to experimental limitations, direct measurements of the acoustic velocity fluctuations were not feasible. The estimates of velocity fluctuation amplitudes obtained here are used in subsequent sections to interpret the observed transition of stable flames from O-SF to O-FREI under forcing, as well as the transition of propagating flames from O-PF to O-FREI.

Equation (5) clearly shows that the velocity forcing experienced by the flame depends on its instantaneous position within the channel due to the spatial variation of the acoustic mode. However, the flame travel distance is small compared to the tube length in all regimes except the propagating flame (PF) and oscillating propagating flame (O-PF) regimes. Consequently, the spatial variation of the velocity field experienced by the flame can be neglected in these regimes (regimes excluding PF and O-PF), and appropriate simplifying assumptions are adopted in the reduced-order models described in the following sections. For the PF and O-PF regimes, this spatial variation is retained. Further details of these assumptions and their validity are provided in Section \ref{sec:fund_freq}.

\section{Forced Response Characteristics} \label{sec:fund_freq}
This section describes the features of forcing different flame regimes at the natural harmonic of the tube. 

\subsection{Flames with Repetitive Extinction and Ignition} \label{subsec:FR}
\begin{figure}
    \centering
    \includegraphics[width=0.85\textwidth]{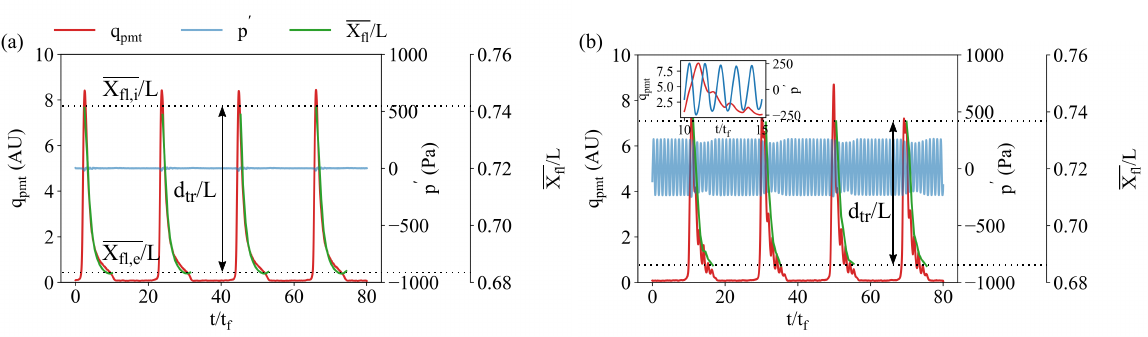}
    \caption{(a) Heat release rate ($q_{pmt}$) of FREI is plotted alongside corresponding pressure fluctuations ($p^{\prime}$) and normalised mean flame position ($\overline{X_{fl}}/L$) under unforced conditions. (b) Corresponding time series plots for imposed upstream pressure fluctuations with an rms amplitude of 197 Pa.}
    \label{fig:FREI_TS}
\end{figure}

Figure \ref{fig:FREI_TS} presents the time series of OH* chemiluminescence ($q_{pmt}$) and the measured upstream pressure fluctuations ($p^{\prime}$) for a flame exhibiting repetitive extinction–ignition behaviour. The corresponding mean flame position ($\overline{X_{fl}}$) is also plotted alongside these signals. Panel (a) presents the baseline, unforced case, while panel (b) corresponds to the forced response at the rms pressure amplitude of 197 Pa.

\begin{figure}
    \centering
    \includegraphics[width=0.85\textwidth]{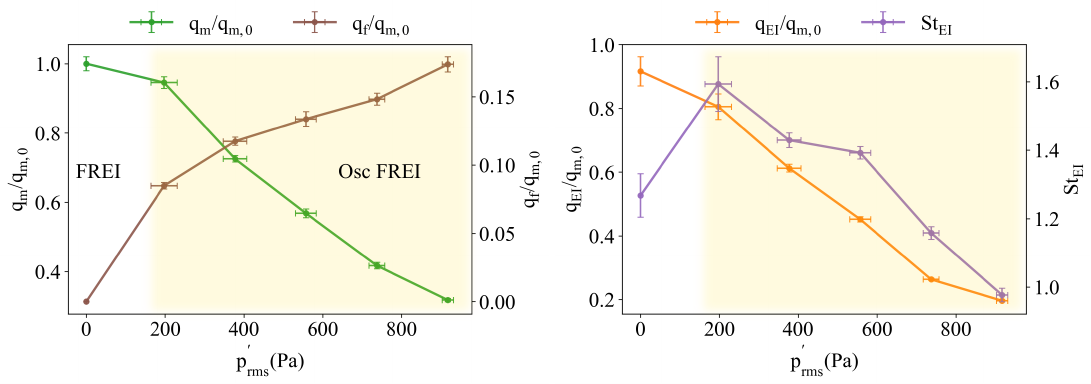}
    \caption{(a) Variation in the mean heat release rate ($q_{m}$) of FREI is plotted against the rms amplitude of the imposed upstream pressure fluctuations. The figure also shows the corresponding variation in heat release rate fluctuations ($q_{f}$) at the forcing frequency. All values are normalised against the mean heat release rate under unforced conditions ($q_{m,0}$). (b) The normalised amplitude of the heat release rate associated with the ignition–extinction cycle ($q_{EI}/q_{m,0}$) is plotted alongside the Strouhal number ($St_{EI}$) characterising this cycle.}
    \label{fig:FREI_Trends}
\end{figure}

\begin{figure}
    \centering
    \includegraphics[width=0.85\textwidth]{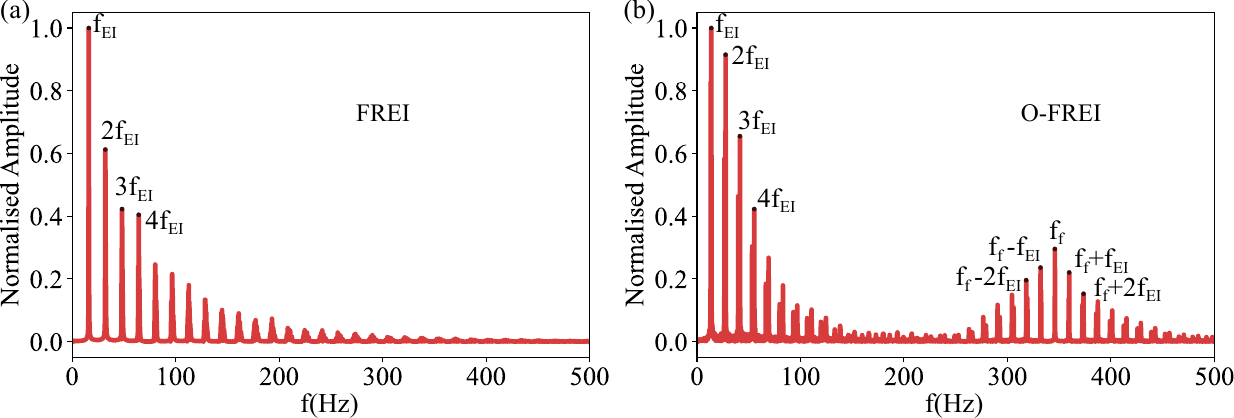}
    \caption{(a) Fast Fourier transform (FFT) amplitude spectrum of the heat release rate signal for an unforced FREI. (b) Corresponding FFT amplitude spectrum for a forced FREI (Oscillating FREI).}
    \label{fig:FREI_NL}
\end{figure}

Unforced FREI ignites at a downstream location close to the external heater ($\overline{X_{fl,i}}/L$ in Fig. \ref{fig:FREI_TS}a), then propagates upstream into a region of progressively lower mean flow temperatures (see Appendix A) and extinguishes after a characteristic travel distance (at $\overline{X_{fl,e}}/L$) due to heat losses to the combustor walls. This behaviour is reflected in the heat release rate plots as a decaying curve following the peak attained after ignition.
The next FREI cycle begins when the unburnt reactants convect downstream, continuously gaining heat from the combustor walls and auto-ignite at the ignition location. The plots in Fig. \ref{fig:FREI_TS}(a) clearly indicate a periodically repeating signal, illustrating the presence of a dominant timescale associated with the ignition–extinction cycle ($t_{EI}$).

When forced at an rms pressure amplitude of 197 Pa (as measured using the microphone M1), FREI starts exhibiting oscillations in heat release rate about the local mean, which is evident from the plots in Fig. \ref{fig:FREI_TS}(b). The qualitative nature of the FREI regime, characterised by auto-ignition, upstream propagation, extinction after a characteristic travel distance, and re-ignition, is retained.
However, external forcing introduces an additional time scale ($t_{f}$) associated with oscillations at the forcing frequency, in addition to the intrinsic ignition–extinction time scale ($t_{EI}$). The plots in Fig. \ref{fig:FREI_TS}(b) correspond to the oscillating FREI (O-FREI) regime illustrated in the flame images presented in Fig. \ref{fig:FREI_Obs}(b).

Fast Fourier Transform (FFT) analysis of the heat release rate signal ($q_{pmt}$) was employed to examine the variations in the amplitudes associated with the different time scales as a function of the upstream forcing pressure in the oscillating FREI (O-FREI) regime. The resulting trends are shown in Fig. \ref{fig:FREI_Trends}.
The mean heat release rate ($q_m$) is found to decrease with increasing forcing pressure, whereas the amplitude associated with the forcing frequency ($q_f$) increases with $p^{\prime}_{rms}$ (Fig. \ref{fig:FREI_Trends}a). Concurrently, the amplitude associated with the ignition–extinction cycle ($q_{EI}$) decreases monotonically as the forcing amplitude is increased (Fig. \ref{fig:FREI_Trends}b). At low forcing pressures, $q_m$ and $q_{EI}$ are significantly larger than $q_f$; however, at the highest forcing pressures explored, all three amplitudes become comparable in magnitude.

An additional parameter used to characterise the ignition–extinction cycle is the Strouhal number associated with the cycle, defined in terms of the ignition–extinction frequency ($f_{EI}$) and the flame travel distance ($d_{tr}$; the distance between the ignition and extinction locations) as:
\begin{gather*}
St_{EI} = \frac{f_{EI} d_{tr}}{\bar{u}}
\tag{6}
\end{gather*}
In the equation, $\bar{u}$ denotes the mean velocity of the premixed reactants. The ignition-extinction Strouhal number, $St_{EI}$, is found to decrease monotonically with increasing $p^{\prime}_{rms}$ (Fig. \ref{fig:FREI_Trends}b).

It is also interesting to note that FFT analysis of the heat release rate signal reveals the emergence of additional spectral components under external forcing. Specifically, the flame exhibits fluctuations at combination frequencies such as $f_f \pm f_{EI}$, $f_f \pm 2f_{EI}$, $f_f \pm 3f_{EI}$, and their higher-order harmonics (Fig. \ref{fig:FREI_NL}b). These components appear in addition to the intrinsic harmonics of the ignition-extinction cycle (e.g., $2f_{EI}$, $3f_{EI}$, and so on), which are present even under unforced conditions (Fig. \ref{fig:FREI_NL}a). 

\begin{figure}
    \centering
    \includegraphics[width=0.85\textwidth]{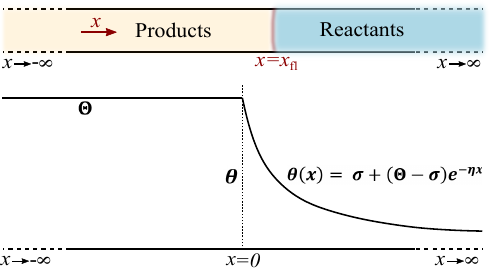}
    \caption{Schematic of the one-dimensional reduction of micro/meso-scale flame dynamics based on the formulation proposed by \cite{minaev_nonlinear_2007}. In the figure, $x$ denotes the position normalised by the flame thickness, and $\theta$ denotes the wall temperature normalised by the adiabatic flame temperature.}
    \label{fig:FREI_Model_Schematic}
\end{figure}

The observed variations in the characteristics of the oscillating FREI regime can be interpreted by augmenting the mesoscale flame model proposed by \cite{minaev_nonlinear_2007} with a harmonic velocity term to account for the effect of upstream forcing. The original model simplifies the dynamics of mesoscale flames in channels to an infinitely extended one-dimensional framework with an imposed wall temperature profile. The thermo-diffusive conservation equations for temperature and reactant concentration are reduced by invoking the large activation energy asymptotic limit, allowing the flame to be treated as a thin discontinuity separating unburnt reactants and burnt products. A schematic illustrating this framework is shown in Fig. \ref{fig:FREI_Model_Schematic}. Here, $x$ denotes the spatial coordinate non-dimensionalised by the flame thickness ($l_{th}$), and $\theta$ denotes the wall temperature normalised by the adiabatic flame temperature ($T_b$). The flame is located at $x = x_{fl}$, with products occupying $x < x_{fl}$ and reactants occupying $x > x_{fl}$, and the domain extends from $-\infty$ to $+\infty$. The parameters $\sigma$ and $\Theta$ represent the normalised unburnt mixture temperature ($T_u/T_b$) and maximum wall temperature ($T_w/T_b$), respectively. The wall temperature profile is assumed to follow an exponential distribution characterised by the decay exponent, $\eta$.

A coordinate transformation, $x = (X_m - X)/l_{th}$, maps the experimental configuration onto the theoretical framework for $x > 0$. The flame solution proposed by \cite{minaev_nonlinear_2007} exists only for $x_{fl} > 0$, corresponding to the region where the wall temperature is described by an exponential profile. This is consistent with our experimental observations, where the flame remains confined between $X = 0$ and $X = X_m$, within which the wall temperature can be approximated by an exponential fit (Fig. \ref{fig:expt_profile}a).

The governing equations are further simplified by assuming quasi-one-dimensional heat transfer (valid for slender channels similar to the present experimental configuration, where transverse gradients can be neglected in comparison with the axial gradients), constant transport properties, and a slowly varying thermal field relative to the flame dynamics. The latter approximation implies that temporal variations in the temperature and concentration fields are small compared to their spatial variations in a frame attached to the flame. Under this assumption, the thermal and concentration fields adjust quasi-instantaneously to the instantaneous flame position, enabling an asymptotic expansion of the governing equations about the steady-state flame solution in terms of the flame-front velocity and acceleration, thereby reducing the problem to a low-dimensional dynamical system \cite{minaev_nonlinear_2007}.

By matching solutions across the preheat and reaction zones and invoking a quasi-steady balance, an analytical relation between the normalised flame temperature ($T_{fl}$) and the effective propagation velocity is obtained, introducing a characteristic logarithmic dependence. The effective propagation velocity is given by $V + \dot{x}_{fl}$, which represents the flame speed relative to the incoming reactant flow velocity ($V$). The resulting closure relation is expressed as:
\begin{align*}
    T_{fl} = & T_V - \frac{2 T^{2}_{fl,s}}{N V^3 Le} \ddot{x}_{fl} \\
    \text{where,} \qquad T_V = & \frac{1}{\left( 1 - \frac{2}{N}ln(V + \dot{x}_f)\right)}
    \tag{7}
\end{align*}
where, $Le$ is the Lewis number of the mixture and $N$ is the normalised activation energy.
Substitution of this relation into the reduced governing equations yields a nonlinear second-order ordinary differential equation describing the evolution of the flame front position:
\begin{align*}
    & K_1 \ddot{x}_{fl} + K_2 \dot{x}_{fl} + K_3 (\dot{x}_{fl})^2  + H T_V + R_1 - R_2 - (1-\sigma)V = 0
    \tag{8}
\end{align*}
where, 
\begin{align*}
    K_1 = & \frac{1-\sigma}{V^2 Le} - \frac{1}{H^3}(VT_{fl,s} +R_1 + R_2) + \frac{2T_{fl,s}^2}{N} \left( \frac{2}{HV} - \frac{H}{V^3 Le} \right) \\
    K_2 = & \frac{1}{H} (VT_V + R_2 + R_1) - (1-\sigma) \\
    K_3 = & \frac{2 \Omega}{H^3} (T_{fl,s} - T_w) \\
    R_1 = & -\Omega e^{(\lambda_2 x_{fl})} \left( \frac{\sigma}{\lambda_2} e^{-\lambda_2 x_{fl}} + \frac{\Theta - \sigma}{\eta + \lambda_2} e^{(-(\eta + \lambda_2) x_{fl})} \right) \\
    R_2 = & \Omega e^{\lambda_1 x_{fl}} \left( -\frac{\sigma}{\lambda_1} e^{-\lambda_1 x_{fl}} -\frac{\Theta - \sigma}{\eta + \lambda_1} e^{-(\eta + \lambda_1)x_{fl}} \right) + \Omega e^{\lambda_1 x_{fl}}\left(-\frac{(\Theta - \sigma)\,\eta}{\lambda_1(\eta + \lambda_1)} \right) \\
    \lambda_1 = & -\frac{V}{2} - \left( \frac{V^2}{4} +\Omega \right)^{1/2} \\
    \lambda_2 = & -\frac{V}{2} + \left( \frac{V^2}{4} +\Omega \right)^{1/2} \\    
    H = & \sqrt{V^2 + 4\Omega} \\
    T_{fl,s} & = \frac{1}{\left( 1 - \frac{2}{N}ln(V)\right)}
    \tag{9}
\end{align*}
Here, $\Omega$ characterises the heat exchange between the combustor walls and the reactant mixture. The readers are referred to the work of \cite{minaev_nonlinear_2007} for further details on the formulation.

\begin{figure}
    \centering
    \includegraphics[width=0.85\textwidth]{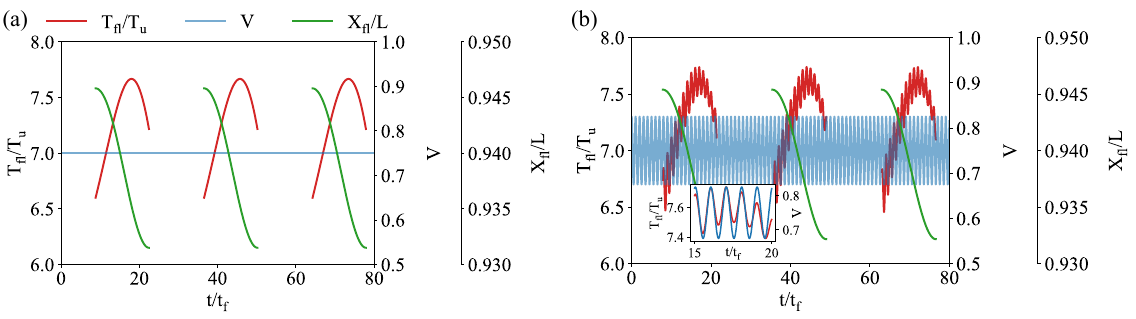}
    \caption{(a) Model-predicted evolution of the normalised flame temperature ($T_{fl}/T_u$) and normalised flame position ($X_{fl}/L$) for the FREI regime under unforced conditions is plotted alongside the normalised reactant mixture velocity ($V$). Panel (b) corresponds to imposed upstream velocity fluctuations of 10\% of the mean flow velocity.}
    \label{fig:FREI_Model}
\end{figure}

The time-series predictions obtained from the above model for unforced FREI, evaluated for parameters corresponding to the present experiments, are shown in Fig. \ref{fig:FREI_Model}(a). For consistency with the experimental measurements, the flame position is expressed in terms of the co-ordinate system $X$. The model evaluates the flame position ($X_{fl}$) and flame temperature ($T_{fl}$) as continuous time series signals, both of which exhibit oscillatory behaviour. It is important to note that flame extinction does not explicitly occur within the model formulation; instead, it is imposed through cut-off limits based on experimental observations. The maximum value of $X$ in the flame position time series corresponds to the closest proximity of the flame to the location of maximum wall temperature and is therefore identified as the ignition point. Conversely, the minimum value of $X$, corresponding to the farthest location from the maximum wall temperature, is interpreted as the extinction point. These cut-off limits make the model-predicted time-series in Fig. \ref{fig:FREI_Model}(a) appear as a series of discontinuous curves (synonymous to the experimental time series presented in Fig. \ref{fig:FREI_TS}a), demonstrating periodic cycles of auto-ignition, upstream propagation, extinction, and re-ignition over a well-defined time interval. The model-predicted flame temperature increase following ignition reaches a peak value and subsequently decreases until extinction, which is qualitatively consistent with the experimental heat release rate variation over an FREI cycle.

It should be noted that the model provides a qualitative description of FREI from a dynamical systems perspective. While the model captures the dynamics (oscillatory nature of the flame) and its response to the imposed thermal gradient, it does not accurately predict quantitative measures such as the ignition-extinction locations, ignition–extinction cycle frequency, or flame travel distance for comparison with the experimental data \cite{minaev_nonlinear_2007}. Nevertheless, the model remains a useful tool for understanding the underlying dynamics of FREI and for interpreting the influence of external forcing in a physically consistent framework.

The effect of upstream forcing is incorporated into the reduced-order model by introducing a harmonic perturbation to the mean flow velocity, such that $V = V_m + V_a \sin(\omega t)$, where $V_m$ is the mean velocity, $V_a$ is the fluctuation amplitude, and $\omega$ is the angular frequency of the imposed velocity fluctuation.
It should be noted that, unlike the experimental configuration—where the magnitude of the acoustic forcing velocity depends on the spatial location of the flame through the mode shape (in addition to the temporal phase), as discussed in Section \ref{sec:vel_est}—this spatial dependence is neglected in the above formulation of $V$. This simplification is justified as the plots in Fig. \ref{fig:FREI_TS} show that the flame remains confined within a narrow spatial region, $0.68 < (X_{fl}/L) < 0.75$. Consequently, the variation in acoustic velocity magnitude arising from the spatial mode shape is negligible and can be neglected.

The modification, $V = V_m + V_a \sin(\omega t)$, remains consistent with the underlying assumptions of the reduced order model provided that the forcing amplitude is sufficiently small. When $V_a/V_m << 1$, the induced variations in the flame-front position and velocity remain moderate, ensuring that the quasi-steady approximation remains valid. Consequently, the thermal and concentration fields continue to adjust quasi-instantaneously to the instantaneous flame position, and the asymptotic expansion in terms of flame-front velocity and acceleration remains applicable.
However, as the forcing amplitude increases, the temporal evolution of the thermal and concentration fields can no longer be treated as a perturbation about the steady flame solution, as the timescale of the flame motion becomes comparable to or faster than the characteristic thermal relaxation timescale. Mathematically, the formulation becomes ill-posed when $V_a \geq V_m$, since the instantaneous velocity $V$ approaches zero or becomes negative (when $\sin (\omega t) = -1$ and $V = V_m - V_a \leq 0$), rendering the logarithmic temperature–velocity relation for $T_{f,s}$ undefined (Equation 9). However, in practice, it is observed that the model becomes increasingly stiff for $V_a > 0.2V_m$, beyond which the solutions are not reliable. Physically, this corresponds to phases of strong convection imposed by the forcing field, which are not adequately captured by the asymptotic model employed. Nevertheless, the qualitative trends predicted by the model at low forcing intensities are governed by robust physical mechanisms and persist at higher forcing amplitudes, although they cannot be quantitatively evaluated within the present reduced-order framework due to the breakdown of its underlying assumptions.

Figure \ref{fig:FREI_Model} (b) illustrates the model time-series of the flame position and flame temperature for forcing amplitudes corresponding to 10\% of the mean flow velocity. The model-predicted flame temperature profile exhibits local oscillations at the forcing frequency superimposed on the ignition–extinction cycles, consistent with the experimental observations in the heat release rate profile shown in Fig. \ref{fig:FREI_TS}(b).

\begin{figure}
    \centering
    \includegraphics[width=0.85\textwidth]{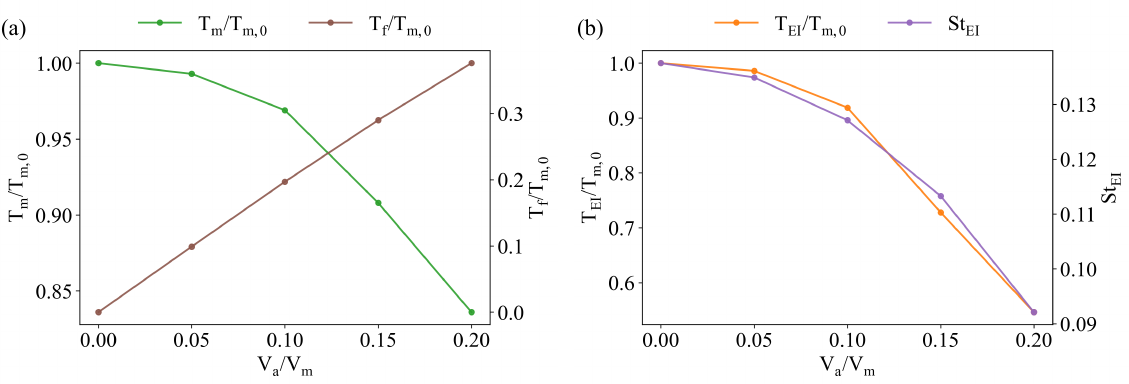}
    \caption{(a) Normalised mean flame temperature ($T_{m}/T_{m,0}$) and normalised fluctuation amplitude at the forcing frequency ($T_{f}/T_{m,0}$) for an oscillating FREI (O-FREI) is plotted against the normalised upstream velocity fluctuation amplitude ($V_a/V_m$). (b) normalised peak ignition–extinction temperature ($T_{EI}/T_{m,0}$) is plotted alongside the Strouhal number characterising this cycle. All temperatures are normalised by the unforced mean temperature.}
    \label{fig:FREI_Model_trends}
\end{figure}

FFT analysis of the model-predicted time series of flame temperature was performed to extract the amplitudes associated with the different time scales present in the signal. The results, shown in Fig. \ref{fig:FREI_Model_trends}(a), indicate that the mean flame temperature ($T_m$) decreases with increasing upstream velocity fluctuations, while the amplitude of the temperature fluctuations at the forcing frequency ($T_f$) increases. These trends are in qualitative agreement with the experimental observations presented in Fig. \ref{fig:FREI_Trends}(a). 
Similarly, the amplitude associated with the ignition-extinction cycle ($T_{EI}$), as well as the Strouhal number characterising the ignition-extinction instability ($St_{EI}$), decreases with increasing upstream velocity fluctuations, consistent with the experimental trends shown in Fig. \ref{fig:FREI_Trends}(b).

\begin{figure}
    \centering
    \includegraphics[width=0.85\textwidth]{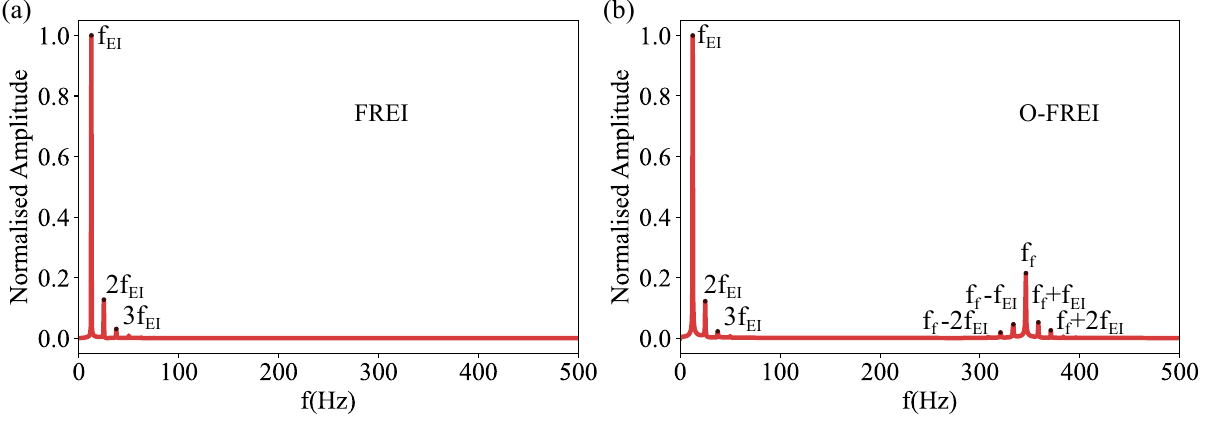}
    \caption{(a) Fast Fourier transform (FFT) amplitude spectrum of the model-predicted flame temperature for an unforced FREI. (b) Corresponding FFT amplitude spectrum for a forced FREI (Oscillating FREI) at $(V_a/V_m) = 10\%$.}
    \label{fig:FREI_model_NL}
\end{figure}

Additionally, the FFT spectrum of the model-predicted flame temperature corresponding to forced FREI (Fig. \ref{fig:FREI_model_NL} b) exhibits components at combination frequencies such as $f_f \pm f_{EI}$ and $f_f \pm 2f_{EI}$, analogous to those observed in the experimental heat release rate spectra. These appear alongside the intrinsic harmonics of the FREI dynamics, including $2f_{EI}$, $3f_{EI}$, and higher-order harmonics (Fig. \ref{fig:FREI_model_NL} a). 

It should be noted that, for the range of upstream forcing pressures explored in this study (197 Pa to 917 Pa), $V_a/V_m$ varies from 2.06 to 9.61, as estimated using the formulation presented in Section \ref{sec:vel_est}. As discussed earlier, the dynamical systems model is not valid within this range. Nevertheless, the trends predicted by the model (Fig. \ref{fig:FREI_Model_trends}) are found to be in qualitative agreement with the experimentally observed trends (Fig. \ref{fig:FREI_Trends}). This indicates that, although the model loses quantitative validity at higher forcing amplitudes, the underlying physical mechanisms governing FREI dynamics remain consistent with those captured by the model.

The observed trends can be understood from the strongly nonlinear dependence of the flame temperature on the upstream flow velocity, as evident from Eqs. (7)–(9). Owing to this nonlinear relation, the thermal response of the flame over a forcing cycle becomes inherently asymmetric. During the high-velocity phase, the increased flow velocity reduces the residence time of the reactant mixture within the heated region, limiting wall-to-gas heat transfer and lowering the preheat temperature of the unburnt mixture. This leads to a pronounced reduction in the flame temperature. In contrast, during the low-velocity phase, the increased residence time promotes heat transfer from the wall, resulting in a rise in flame temperature. However, this recovery is constrained by the finite local wall temperature and delayed by the thermal inertia of the system. Consequently, the reduction in flame temperature during the high-velocity phase exceeds the corresponding increase during the low-velocity phase, leading to a net decrease in the mean flame temperature and a reduction in the peak temperature attained during the ignition–extinction cycle. 
A rigorous mathematical formulation for estimating the cycle-averaged flame temperature, highlighting the effect of asymmetric thermal loading, is presented in Section \ref{subsec:SF} for oscillating stable flames. Such a direct analytical estimation is not feasible for oscillating FREI, as the flame temperature and flame position are intrinsically coupled through the nonlinear dynamical system described by Eqs. (7)–(9), precluding a closed-form evaluation over a forcing cycle.

The reduction in effective flame temperature directly influences the spatial dynamics of the flame. Lower temperatures during the high-velocity phase promote earlier extinction, causing the flame to extinguish closer to the heated region. Conversely, the partial recovery in temperature during the low-velocity phase triggers ignition before the flame reaches its unforced downstream ignition location. 
As a result, the axial travel distance of the flame reduces, thereby reducing the corresponding Strouhal number.

Additionally, the model also incorporates a direct coupling between flame acceleration and temperature through the inertial term $\left(\frac{2 T^{2}_{fl,s}}{N V^3 Le} \ddot{x_{fl}}\right)$ in Eq. (7). This term introduces a fast response mechanism, whereby velocity fluctuations induce flame acceleration, which in turn modulates the flame temperature at the forcing frequency. As the forcing amplitude increases, this coupling becomes stronger, leading to enhanced temperature oscillations. Therefore, while the mean flame temperature decreases, the amplitude of temperature fluctuations increases with forcing, consistent with the experimental observations.

\subsection{Stable flames} \label{subsec:SF}

\begin{figure}
    \centering
    \includegraphics[width=0.85\textwidth]{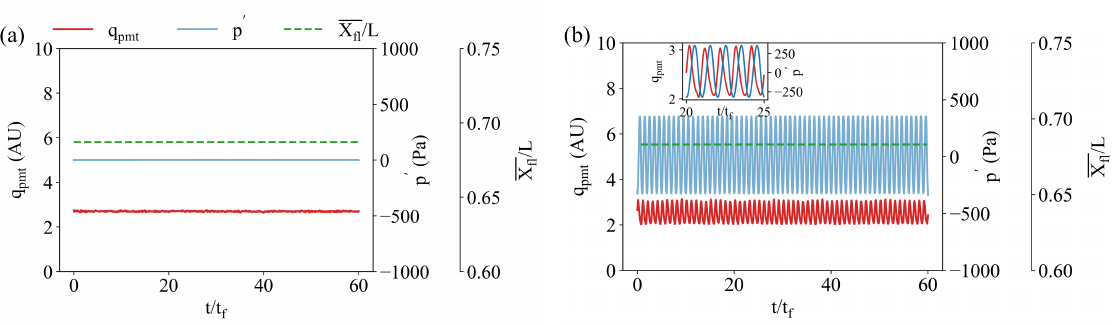}
    \caption{(a) Heat release rate ($q_{pmt}$) of unforced stable flames is plotted alongside the measured pressure fluctuations ($p^{\prime}_{rms}$) and normalised mean flame position ($\overline{X_{fl}}/L$). (b) Corresponding time series plots for imposed upstream pressure fluctuations with an rms amplitude of 197 Pa (oscillating stable flame).}
    \label{fig:SF_TS}
\end{figure}

Figure \ref{fig:SF_TS}(b) presents the time-series evolution of the OH* chemiluminescence ($q_{pmt}$), the mean flame position ($\overline{X_{fl}}$), and the imposed pressure fluctuations ($p^{\prime}$) measured by microphone M1 for a stable flame subjected to external forcing ($p^{\prime}_{rms} \approx 197$ Pa). The plots corresponding to the unforced stable flame under the same operating conditions are shown in Fig. \ref{fig:SF_TS}(a), where the flame is stabilised at an axial location ($X_{fl,s}$) in the combustor tube determined by the imposed wall temperature profile. The heat release rate of an unforced stable flame remains at a near-constant value (Fig. \ref{fig:SF_TS}a).

As discussed in Section \ref{sec:glo_obs}, at low forcing amplitudes ($p^{\prime}_{rms} \approx 197$ Pa), the flame oscillates about its mean position. This behaviour is evident from the flame images shown in Fig. \ref{fig:SF_Obs}(b) and from the OH* chemiluminescence signal in Fig. \ref{fig:SF_TS}(b), which illustrates oscillations in heat release rate about the local mean. This response corresponds to the oscillating stable flame (O-SF) regime.

\begin{figure}
    \centering
    \includegraphics[width=0.85\textwidth]{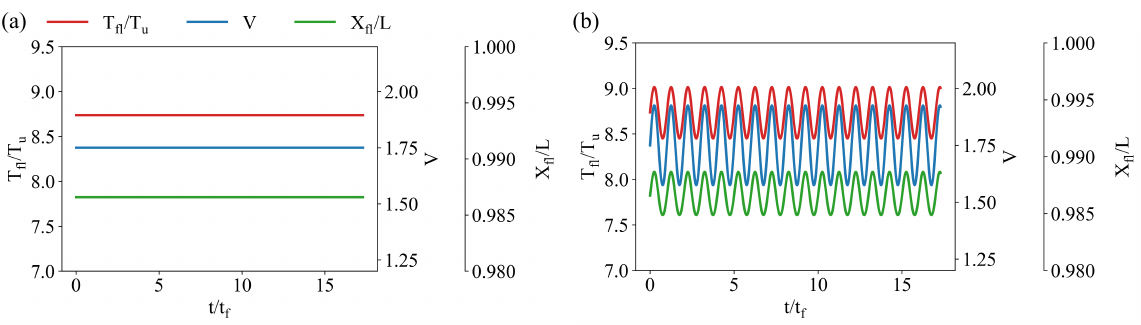}
    \caption{(a) Model-predicted evolution of the normalised flame temperature ($T_{fl}/T_{u}$) and normalised flame position (${X}_{fl}/L$) for stable flames under unforced conditions, plotted alongside the reactant mixture velocity. Panel (b) corresponds to imposed upstream velocity fluctuations of 10\% of the mean flow velocity.}
    \label{fig:Model_SF}
\end{figure}

The O-SF regime can be modelled by extending the framework proposed by \cite{minaev_nonlinear_2007} for stable flames in micro and mesoscale channels. The assumptions and simplifications underlying the reduced-order model are discussed in detail in Section \ref{subsec:FR}. 
For an unforced stable flame, the flame temperature ($T_{fl}$) can be expressed as,
\begin{align*}
    T_{fl} = T_{fl,s} = \frac{1}{\left( 1 - \frac{2}{N}ln(V)\right)}
    \tag {10}
\end{align*}
The subscript, $s$, in the above equation represents stable flames. Following the evaluation of the flame temperature, flame position ($x_{fl}$) can then be estimated using the implicit equation: 
\begin{align*}
    H T_{fl} + R_1 (x_{fl}) - R_2 (x_{fl}) - (1-\sigma)V = 0
    \tag {11}
\end{align*}
It should be noted that Eqs. (10) and (11), which corresponds to stable flames, can be directly obtained by substituting $\dot{x}_{fl} = 0$ and $\ddot{x}_{fl} = 0$ into the reduced-order model for FREI presented in Eqs. (7) and (8). This shows that the stable flame solution corresponds to the fixed point of the reduced-order ODE for FREI given in Eq. (8). This is consistent with the formulation of the ODE, which is derived from an asymptotic expansion of the governing equations about the steady-state solution. It should be noted that, although the analytical model for stable flames proposed by \cite{minaev_nonlinear_2007} captures the qualitative flame dynamics, direct quantitative comparison of flame position, heat release rate, or flame temperature with experimental observations is not feasible.

Following the approach described in Section \ref{subsec:FR}, the effect of upstream forcing is incorporated by introducing a harmonic component to the reactant mixture velocity, such that $V = V_m + V_a \sin(\omega t)$.
It should be noted that, analogous to the modelling assumption in Section \ref{subsec:FR}, the spatial dependence of the acoustic velocity perturbation amplitude is neglected. This is justified because the oscillating stable flame is localised within a narrow region centred around $X_{fl,s}$, and therefore the variation in acoustic velocity due to the mode shape is not expected to be significant.

The time-series responses predicted by the model for both unforced and forced conditions are shown in Fig. \ref{fig:Model_SF}. Unforced stable flames exhibit a steady flame temperature with no flame-front motion (Fig. \ref{fig:Model_SF}a). Qualitatively consistent with the experimental observations presented in Fig. \ref{fig:SF_TS}(b), forcing induces oscillations in the flame temperature about its mean value, accompanied by back-and-forth motion of the flame front about the mean axial position (Fig. \ref{fig:Model_SF}b).

\begin{figure}
    \centering
    \includegraphics[width=0.85\textwidth]{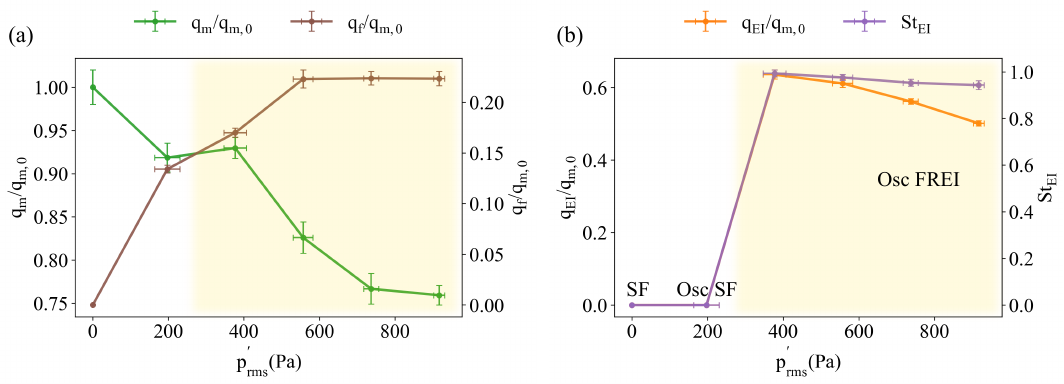}
    \caption{(a) Normalised mean heat release rate ($q_m/q_{m,0}$) of SF is plotted against the rms amplitude of the imposed upstream pressure fluctuations. The panel also plots the variation in the normalised heat release rate fluctuations ($q_f/q_{m,0}$) at the forcing frequency. (b) The amplitude of the normalised heat release rate associated with the ignition–extinction cycle ($q_{EI}/q_{m,0}$) is plotted alongside the Strouhal number characterising this cycle.}
    \label{fig:SF_Trends}
\end{figure}

Experimental plots in Fig. \ref{fig:SF_Trends}(a) show that, in the oscillating stable flame (O-SF) regime (unshaded region), the mean heat release rate ($q_m$) drops when forced. Assuming analogous qualitative trends between the experimentally measured heat release rate and the model-predicted flame temperature, the observed reduction in heat release rate can be attributed to asymmetric thermal loading over a forcing cycle. This behaviour can be understood from a mathematical perspective by estimating the cycle-averaged flame temperature ($T_{m}$). By averaging Eq. (10) over a forcing cycle, an expression for the cycle-averaged flame temperature ($T_m$) can be obtained. Prior to averaging, the following simplification is introduced by expressing $V$ as $V_m + v$, where $v = V_a \sin(\omega t)$, and evaluating the expression in the limit $V_a \ll V_m$ (simplifying assumption to derive an analytical expression). 
\begin{align*}
    T_{fl,s} = \frac{1}{\left( 1 - \frac{2}{N}ln(V_m + V_a \sin(\omega t))\right)} = \frac{1}{\left( 1 - \frac{2}{N} \left[ ln(V_m) + \frac{v}{V_m} - \frac{v^2}{2V_m^2} + O(v^3) \right] \right)}
    \tag {12}
\end{align*}
Defining $k=2/N$ and $p = 1 - (k \space ln(V_m))$, the above expression can be simplified as,
\begin{align*}
    T_{fl,s} = \frac{1}{p} \left( 1 + \frac{kv}{pV_m} + \left( \frac{k^2}{p^2V_m^2} - \frac{k}{2pV_m^2} \right)v^2 +O(v^3) \right)
    \tag {13}
\end{align*}
The mean flame temperature over a forcing cycle is then obtained by time-averaging Eq. (13) over one forcing period ($t_f$).
\begin{align*}
    T_m = & \frac{1}{t_f} \int_{0}^{t_f} T_{fl,s} dt \\
    T_m = & \frac{1}{t_f}\left(\frac{2\pi}{p}\right) +  \frac{1}{t_f} \left(\frac{\pi V_a^2 k}{p^2V_m^2}\right) \left( \frac{k}{p} - \frac{1}{2} \right)
    \tag {14}
\end{align*}
The first term in the above equation is independent of $V_a$ and represents the cycle-averaged flame temperature under unforced conditions. The second term is negative since $k/p < 0.5$, corresponding to the conditions of the study. Therefore, $T_m(V_a) < T_m(V_a=0)$, which implies that the cycle-average flame temperature drops when forced. This is synonymous with the trends observed in the O-FREI model in Section \ref{subsec:FR}, wherein we find the mean flame temperature drops with increasing forcing amplitudes. 

\begin{figure}
    \centering
    \includegraphics[width=0.85\textwidth]{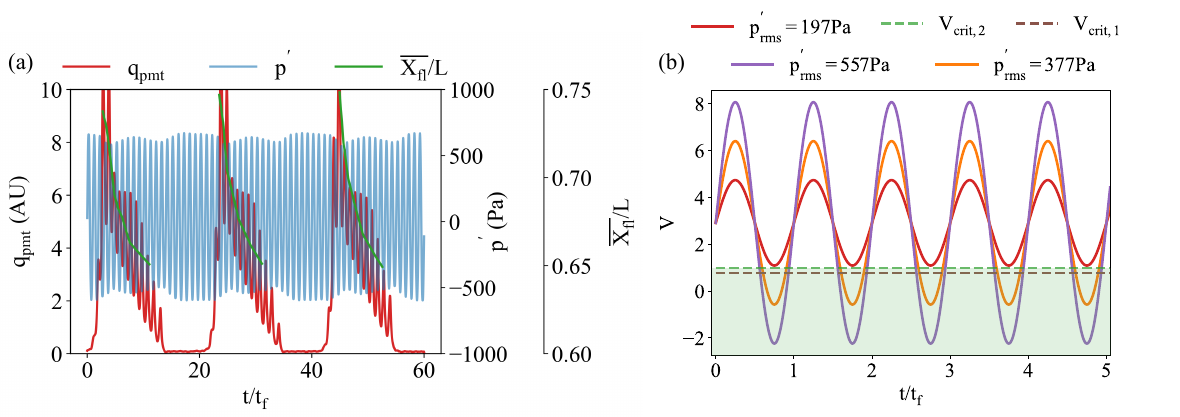}
    \caption{(a) Heat release rate ($q_{pmt}$) of a forced stable flame at an rms pressure amplitude of 377 Pa, plotted alongside the measured pressure signal and the normalised mean flame position ($\overline{X_{fl}}/L$). (b) Time-series variation of the normalised upstream velocity ($V$) corresponding to forcing rms pressure amplitudes of 197 Pa, 377 Pa, and 557 Pa, respectively. The green and brown dotted lines indicate the upper and lower limits of $V$ within which the stable flame solution becomes unstable.}
    \label{fig:SF_Limits}
\end{figure}

When stable flames are forced at $p^{\prime}_{rms} \geq 377$ Pa, the flame transitions to the oscillating FREI (O-FREI) regime. The experimental time-series data shown in Fig. \ref{fig:SF_Limits}(a), along with the corresponding flame images in Fig. \ref{fig:SF_Obs}(c), illustrate the O-FREI response when stable flames are forced at $p^{\prime}_{rms} \sim 377$ Pa. 

\begin{figure}
    \centering
    \includegraphics[width=0.85\textwidth]{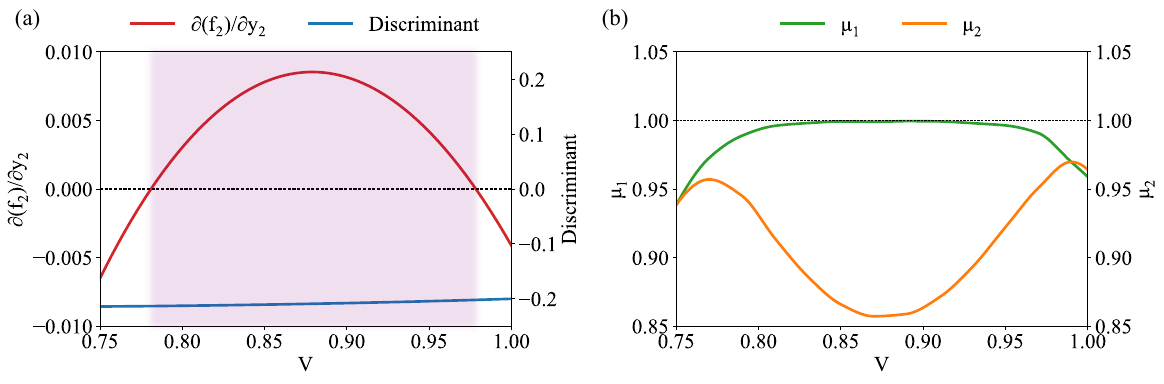}
    \caption{(a) Variation of $\partial f_2/\partial y_2$ and the discriminant associated with the system of ODEs describing the FREI dynamical system, evaluated at the steady flame solution, as a function of the normalised mixture velocity ($V$). (b) Magnitude of the Floquet multipliers corresponding to the O-FREI limit cycle, plotted as a function of $V$.}
    \label{fig:SF_Transition_P2}
\end{figure}

The transition from oscillating stable flame behaviour to O-FREI behaviour at higher forcing intensities can be understood by evaluating the stability of the stable flame solution across different values of the normalised mixture velocity ($V$). As discussed earlier, the reduced-order ODE for FREI presented in Eq. (8) is derived from an asymptotic expansion of the governing equations about the stable flame solution, and its fixed point corresponds to the stable flame solution (Eq. 11).
The ODE in Eq. (8) can be expressed as a system of two first-order ODEs in $y_1$ and $y_2$, where $y_1 = x_{fl}$ and $y_2 = \dot{x}_{fl}$, as,
\begin{align*}
    \dot{y_1} = f_1 = & y_2 \\
    \dot{y_2} = f_2 = & \frac{1}{K_1}\left[- K_2 \dot{x}_{fl} - K_3 (\dot{x}_{fl})^2  - H T_V - R_1 + R_2 + (1-\sigma)V \right]
    \tag {15}
\end{align*}
By evaluating the Jacobian of the system of first-order ODEs about the fixed point, the stability of the stable flame solution can be determined. The Jacobian matrix for the system is of the form,
\begin{align*}
    \renewcommand{\arraystretch}{1.5} 
    \begin{bmatrix}
    \frac{\partial f_1}{\partial y_1} & \frac{\partial f_1}{\partial y_2} \\
    \frac{\partial f_2}{\partial y_1} & \frac{\partial f_2}{\partial y_2}
    \end{bmatrix}
    =
    \renewcommand{\arraystretch}{1.5} 
    \begin{bmatrix}
    0 & 1 \\
    \frac{\partial f_2}{\partial y_1} & \frac{\partial f_2}{\partial y_2}
    \end{bmatrix}
\tag{16}
\end{align*}
The eigenvalues of the Jacobian matrix can be expressed as
\begin{align*}
    \Lambda = \frac{1}{2} \left[ \frac{\partial f_2}{\partial y_2} \pm \sqrt{\left(\frac{\partial f_2}{\partial y_2}\right)^2 + 4 \frac{\partial f_2}{\partial y_1}} \right]
    \tag{17}.
\end{align*}
The discriminant of $\Lambda$ is negative over the entire range of $V$, as shown in Fig. \ref{fig:SF_Transition_P2}(a). The real part of $\Lambda$, which represents the exponential amplitude growth rate of perturbations in time, depends only on $({\partial f_2}/{\partial y_2})$. Evaluating this term at the stable flame solution $(x_{fl,s}, 0)$ shows that, in the normalised velocity range $0.78 \leq V \leq 0.978$, $\left.({\partial f_2}/{\partial y_2})\right|_{(x_{fl,s}, 0)} > 0$, which implies that the stable flame solution becomes unstable in this range (Fig. \ref{fig:SF_Transition_P2}a). Outside this range, the stable flame solution is an attractor.
Similar critical velocity ranges ($0.78 \leq V \leq 0.978$) were reported by \cite{maruta_characteristics_2004} and \cite{minaev_nonlinear_2007}, wherein the stable flame loses its stability within a finite velocity range, and the system exhibits FREI dynamics corresponding to a limit cycle.

It is hypothesised that, during a forcing cycle of a stable flame, when acoustic forcing drives $V(=V_m + V_a \sin(\omega t))$ below the upper critical limit, $V=0.978$, the instantaneous governing equations belong to the parameter regime where the stable flame solution does not exist as an attractor. Consequently, the system enters the basin of attraction of the FREI solution, and the flame transitions from O-SF to the O-FREI regime.
An estimate of the Floquet multipliers for the O-FREI limit cycle further supports this argument. Analogous to using the Jacobian to evaluate the stability of a stationary fixed point, the stability of the O-FREI limit cycle can be assessed by evaluating the Floquet multipliers. These correspond to the eigenvalues of the state transition matrix ($\Psi$), which governs the evolution of perturbations with respect to the limit cycle over one period of oscillation.
For a two-dimensional system, the multipliers correspond to an along-orbit multiplier ($\mu_1$) and a transverse multiplier ($\mu_2$). The along-orbit multiplier characterises the evolution of perturbations along the trajectory, while the transverse multiplier characterises the growth or decay of perturbations normal to the trajectory.
The state transition matrix ($\Psi$) is computed by augmenting the system of ODEs (Eq. 15) with the variational equation $d\Psi/dt = J(t) \Psi$ and $\Psi(0)=I$, where $I$ is the identity matrix and $J(t)$ is the Jacobian of the system of ODEs evaluated along the settled O-FREI orbit at each instant. The resulting system is then integrated over one period ($t_{EI}$) to obtain $\Psi$. The readers are referred to the work of \cite{parker2012practical} for details on the numerical evaluation of the Floquet multipliers. 
For the present configuration, in the normalised velocity range $0.78 \leq V \leq 0.978$, the Floquet multipliers have their magnitudes less than unity, indicating that the perturbations decay over time and implying that the O-FREI solution corresponds to a stable limit cycle (Fig. \ref{fig:SF_Transition_P2}b). 

Consistent with the above hypothesis, Fig. \ref{fig:SF_Limits}(b) shows that, when forced at an rms pressure amplitude of $197$ Pa, the corresponding instantaneous value of $V$ experienced by the flame (estimated using Eq. (5) in Section \ref{sec:vel_est}) remains within the range of $V$ for which the stable flame solution exists ($V>0.978$). Consequently, the flame does not transition to O-FREI. 
However, as the forcing rms pressure amplitude increases to $377$ Pa and above, the corresponding value of $V$ drops below the critical limit ($V<0.978$), causing the flame to transition to the O-FREI regime.

Analogous to the experimental observations and model predictions discussed in Section \ref{subsec:FR} for the O-FREI regime, following the transition to O-FREI, the mean heat release rate ($q_{m}$ in Fig. \ref{fig:SF_Trends}a), the amplitude of the heat release rate associated with the ignition–extinction cycle ($q_{EI}$ in Fig. \ref{fig:SF_Trends}b), and the Strouhal number characterising the ignition–extinction cycle ($St_{EI}$ in Fig. \ref{fig:SF_Trends}b), all decrease with increasing forcing intensity. However, unlike the trends observed in Section \ref{subsec:FR}, the amplitude of the heat release rate fluctuations at the forcing frequency is found to saturate and remain at a near constant value with increasing $p^{\prime}_{rms}$ ($q_{f}$ in Fig. \ref{fig:SF_Trends}a).
This behaviour is a characteristic of the dynamical model described by Equations (7)-(9) for the O-FREI regime. At low forcing intensities, velocity fluctuations directly drive flame-front acceleration ($\ddot{x_{fl}}$), which modulates flame temperature through the inertial term $\left(\frac{2 T^{2}_{fl,s}}{N V^3 Le} \ddot{x_{fl}}\right)$ in Equation (7). As the forcing intensity increases (i.e., as $V_a$ increases), nonlinear damping ($K_3 (\dot{x}_{fl})^2$) in the flame-front dynamics (Eq. 9) becomes significant, limiting further growth of flame-front velocity and, consequently, the acceleration response. This results in the saturation of the flame temperature and heat-release-rate oscillations at higher forcing amplitudes, as observed in Fig. \ref{fig:SF_Trends}(a).

It is further noted that, within the forcing pressure range explored, forcing a FREI flame does not lead to saturation of the heat-release-rate oscillations, whereas such saturation is observed when stable flames are forced. This difference arises from changes in the coefficients of the dynamical model resulting from the distinct operating conditions associated with the FREI and SF regimes. In particular, the higher mean flow velocity ($V_m$) of stable flames, together with the higher activation energy ($N$) under fuel-lean conditions, reduces the magnitude of the inertial coefficient in Equation (7), thereby weakening temperature modulation due to flame-front acceleration. As a result, nonlinear saturation of temperature and heat-release-rate fluctuations occurs at lower forcing thresholds in the SF regime compared to the FREI regime.

\subsection{Propagating flames} \label{subsec:PF}

\begin{figure}
    \centering
    \includegraphics[width=0.85\textwidth]{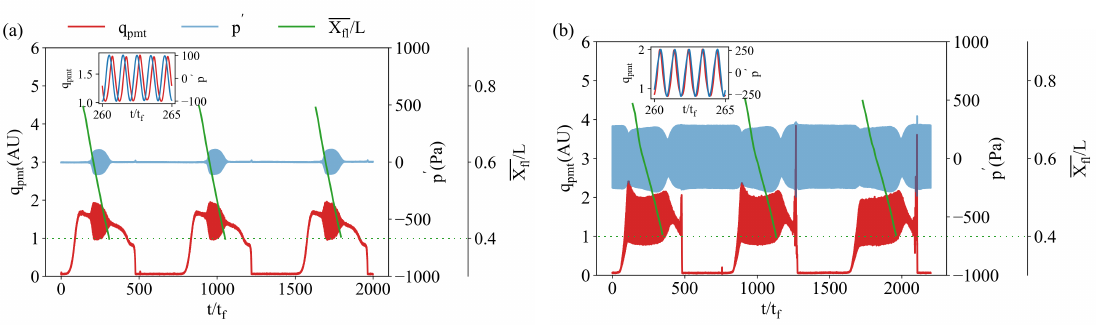}
    \caption{(a) Heat release rate ($q_{pmt}$) of an unforced propagating flame is plotted alongside the measured pressure fluctuations ($p^{\prime}_{rms}$) and normalised mean flame position ($\overline{X_{fl}}/L$). (b) Corresponding time series plots for imposed upstream pressure fluctuations with an rms amplitude of 197 Pa (oscillating propagating flame regime).}
    \label{fig:PF_TS}
\end{figure}

Figure \ref{fig:PF_TS} shows the time-series evolution of the heat release rate, mean flame position, and measured pressure fluctuations for a propagating flame under unforced and forced conditions in panels (a) and (b), respectively.
As evident from Fig. \ref{fig:PF_TS}(a), the unforced propagating flame, upon ignition, travels upstream and reaches the upstream end of the combustor tube while exhibiting thermoacoustic coupling during the propagation phase. This coupling is highlighted in the inset plot of Fig. \ref{fig:PF_TS}(a), which shows a strong coherence between the heat release rate and pressure fluctuations, with a minor phase offset between the two signals. 
It should be noted that the pressure signal shown in the figure has been temporally shifted to account for the acoustic time delay between the microphone location and the instantaneous flame position. 
Additionally, the plots of the mean flame position in Fig. \ref{fig:PF_TS}(a,b) are truncated at an axial distance of $\overline{X_{fl}/L} \sim 0.4$, as the imaging frame was restricted to $\overline{X_{fl}/L} \geq 0.4$ due to practical constraints of the experimental setup. 

Our previous study \cite{aravind_dynamics_2025} established that the frequency of the coupled oscillations in heat release rate and pressure corresponds to the first harmonic of the combustor tube and can be predicted using frequency-domain analysis of the combustor tube \cite{aravind_dynamics_2025}.
Thermoacoustic oscillations during the propagation phase of the flame were found to induce back-and-forth motion of the flame front \cite{aravind_dynamics_2025}, enhancing mixing between reactants and products across the flame front. This mechanism was proposed to sustain flame propagation till the upstream end of the tube without extinction, even at an equivalence ratio of 0.8. In contrast, regimes such as FREI, observed near stoichiometric conditions, did not develop thermoacoustic oscillations during the propagation phase and were found to extinguish after a characteristic travel distance. 

Additionally, the slope of the flame trajectories in Fig. \ref{fig:PF_TS}(a) indicates that the flame displacement rate $(dX_{fl}/dt)$ remains nearly constant for $(X_{fl}/L) < 0.65$. The flame displacement rate can be expressed as the difference between the velocity of the unburnt reactants entering the flame front ($\bar{u}$) and the laminar flame speed ($S_L$):
\begin{align*}
\left(\frac{dX_{fl}}{dt}\right) = \bar{u} - S_L
\tag{18}
\end{align*}
For $(X_{fl}/L) < 0.65$, the imposed wall temperature profile is close to the upstream mixture temperature (Fig. \ref{fig:expt_profile}a). 
However, as the mixture propagates downstream (prior to ignition) beyond this region ($(X_{fl}/L) > 0.65$), it progressively gains heat from the combustor walls, resulting in an increase in its mean temperature (Appendix A). Since the combustor tube is open at $X = L$, the pressure rise associated with this temperature increase is negligible and manifests instead as a reduction in the density of the reactant mixture with increasing $X$.
For the conditions considered in this study, where the incoming mass flow rate of the reactants is maintained constant, a decrease in density along $X$ leads to a corresponding increase in the bulk flow velocity of the reactants ($\bar{u}$). Concurrently, following the increase in mean flow temperature with $X$, the laminar flame speed of the mixture ($S_L$) is also expected to increase. Consequently, for $(X_{fl}/L) > 0.65$, the flame displacement rate $(dX_{fl}/dt)$ depends on the spatial variations of $\bar{u}(X)$ and $S_L(X)$. However, for $(X_{fl}/L) < 0.65$, both quantities can be treated as approximately constant since the imposed wall temperature in this region is close to the upstream mixture temperature and therefore $(dX_{fl}/dt)$ remains nearly constant, consistent with the experimental observations in Fig. \ref{fig:PF_TS}(a).

When the propagating flame is externally forced at its thermoacoustic coupling frequency, the amplitude of the thermoacoustic oscillations increases, as evident from the inset plot in Fig. \ref{fig:PF_TS}(b). The plots further show that, with external forcing, the phase difference between the heat release rate and pressure fluctuations decreases, and the coherence between the two signals increases.
The flame images corresponding to the time-series plots in Fig. \ref{fig:PF_TS}(b) are shown in Fig. \ref{fig:PF_Obs}(b). In this regime (oscillating propagating flame regime), the qualitative characteristics of the propagating flame are preserved, with the flame traversing the full length of the combustor tube while exhibiting strong coupled oscillations in heat release rate and pressure. This flame behaviour can be modelled using a coupled framework that integrates acoustic time-domain analysis with a flame propagation model.

\begin{figure}
    \centering
    \includegraphics[width=0.85\textwidth]{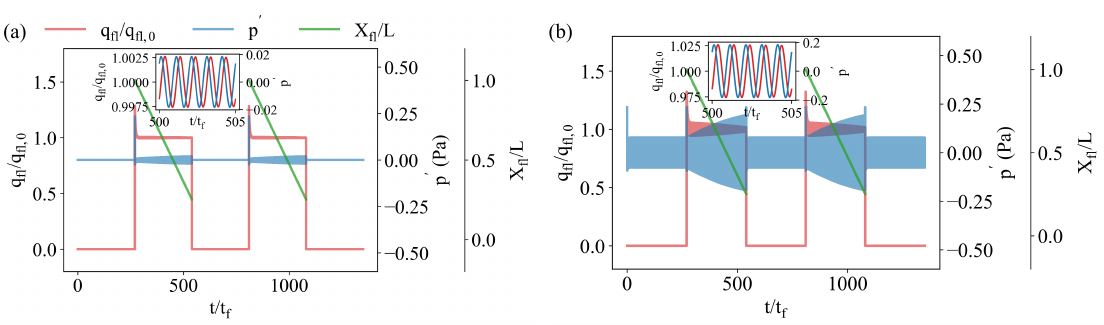}
    \caption{(a) Model-predicted evolution of the normalised heat release rate ($q_{fl}/q_{fl,0}$) and the corresponding pressure fluctuations ($p^{\prime}_{rms}$) for propagating flames under unforced conditions, plotted alongside the normalised mean flame position ($\overline{X_{fl}}/L$). Panel (b) corresponds to forced conditions with the normalised forcing amplitude ($A_e$) of 0.1. Here, $q_{fl,0}$ is the baseline heat release rate corresponding to the fuel consumption rate.}
    \label{fig:Model_PF}
\end{figure}

\begin{figure}
    \centering
    \includegraphics[width=0.85\textwidth]{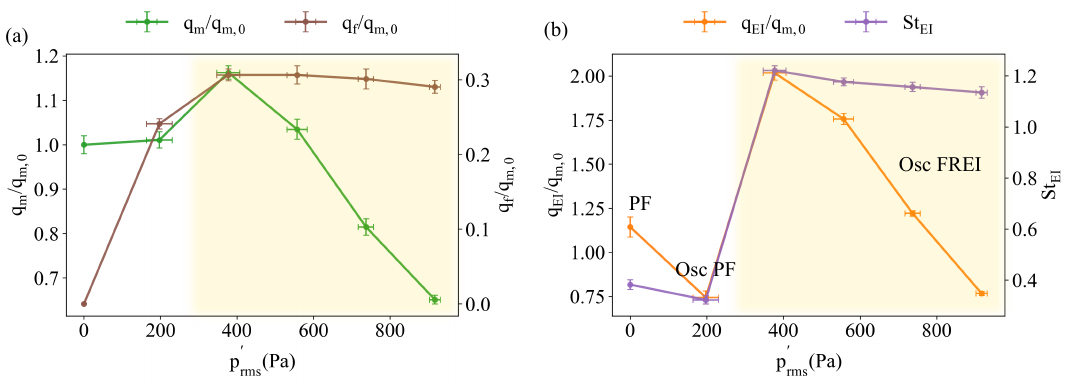}
    \caption{(a) Normalised mean heat release rate ($q_m/q_{m,0}$) of PF is plotted against the rms amplitude of the imposed upstream pressure fluctuations. The panel also plots the variation in the normalised heat release rate fluctuations ($q_f/q_{m,0}$) at the forcing frequency. (b) The amplitude of the normalised heat release rate associated with the ignition–extinction cycle ($q_{EI}/q_{m,0}$) is plotted alongside the Strouhal number characterising this cycle.}
    \label{fig:PF_Trends}
\end{figure}

To account for thermoacoustic coupling in the propagating flame regime, the acoustic governing equations are simplified using a Galerkin-based approach \cite{lieuwen_unsteady_2012, culick_combustion_1996}. In this formulation, the acoustic partial differential equations are reduced to a set of time-domain ordinary differential equations by expanding the acoustic field in terms of the duct eigenmodes and projecting the governing equations onto these modes. As a result, each acoustic mode behaves as a damped harmonic oscillator forced by unsteady heat release rate and external forcing (analogous to upstream forcing in the current study). The unsteady heat release rate is accounted for using a cubic heat release rate model \cite{noiray2008unified, mohan_self-excited_2023}. The combustor domain is simplified by assuming an acoustically closed boundary at $X=0$ (a simplifying approximation motivated by the strong impedance mismatch at the boundary) and an acoustically open boundary at $X=L$, corresponding to the open atmospheric end. The governing ordinary differential equation for the first (dominant) acoustic mode of the combustor tube can be expressed as,
\begin{align*}
    \ddot{\eta} + k^2 \eta + 2 \epsilon \omega \dot{\eta} = \frac{k Z}{\gamma M} \sin \left(\frac{\pi X_f}{L} \right) \eta + \beta_1 \dot{\eta} - \beta_2 \dot{\eta} \eta + F_e
    \tag{19}
\end{align*}
Here, the acoustic velocity fluctuations are expressed as $u^{\prime} = \eta \sin(\pi X_f / 2L)$, ensuring that the boundary conditions at the ends of the combustor tube are satisfied. $k$ denotes the non-dimensional wavenumber, $\omega$ is the non-dimensional angular frequency and $\epsilon$ represents the acoustic damping coefficient. The term $\left( \frac{k Z}{\gamma M} \sin (\pi X_f) \eta + \beta_1 \dot{\eta} - \beta_2 \dot{\eta} \eta \right)$ denotes unsteady heat release rate feeding energy to the acoustic field. Here, $Z$ characterises the coupling strength between the acoustic velocity fluctuations and the unsteady heat release rate, $M$ is the Mach number of the bulk flow, and $\beta_1$ and $\beta_2$ are the coefficients of the cubic heat release rate model \cite{mohan_self-excited_2023}. 
The term $F_e (=A_e \sin(\omega t))$ accounts for the effect of external forcing and is zero for an unforced propagating flame.

The above ODE is coupled with the flame propagation model described in Eq. (18) to account for the upstream motion of the flame. The coupled model (Eqs. 18 and 19) captures upstream flame propagation along with thermoacoustic oscillations during the propagation phase.
Initially, the flame is assumed to ignite at a prescribed axial location ($\overline{X_{fl,i}}$), which is provided as an input to the model. The acoustic field is then evaluated using the Galerkin-based formulation (Equation 19), and the flame position is subsequently updated using the flame propagation model (Equation 18) in an iterative manner until the flame reaches the upstream end of the tube ($X = 0$), at which point extinction is imposed. The next cycle begins after a time interval of $\overline{X_{fl,i}}/\bar{u}$, corresponding to the time required for the reactant mixture to convect downstream (gaining heat from the combustor walls), and auto-ignite at $\overline{X_{fl,i}}$; repeating the propagating flame cycle \cite{aravind_dynamics_2025}. 
The time-series predictions from the model, depicting the evolution of the flame, its unsteady heat release rate and associated pressure fluctuations, are shown in Fig. \ref{fig:Model_PF}. Consistent with the experimental observations, the unforced flame exhibits coupled heat release rate - pressure oscillations during the flame propagation phase (Fig. \ref{fig:Model_PF}a). Fig. \ref{fig:Model_PF}(b) further demonstrates that external forcing (implemented through the term $F_e$ in Eq. 19) amplifies these oscillations in the heat release rate and pressure. This behaviour is in agreement with the experimentally observed dynamics in the oscillating propagating flame (O-PF) regime in Fig. \ref{fig:PF_Trends}a (unshaded region), where $q_f$ increases with $p^{\prime}_{rms}$ in the O-PF regime.

\begin{figure}
    \centering
    \includegraphics[width=0.85\textwidth]{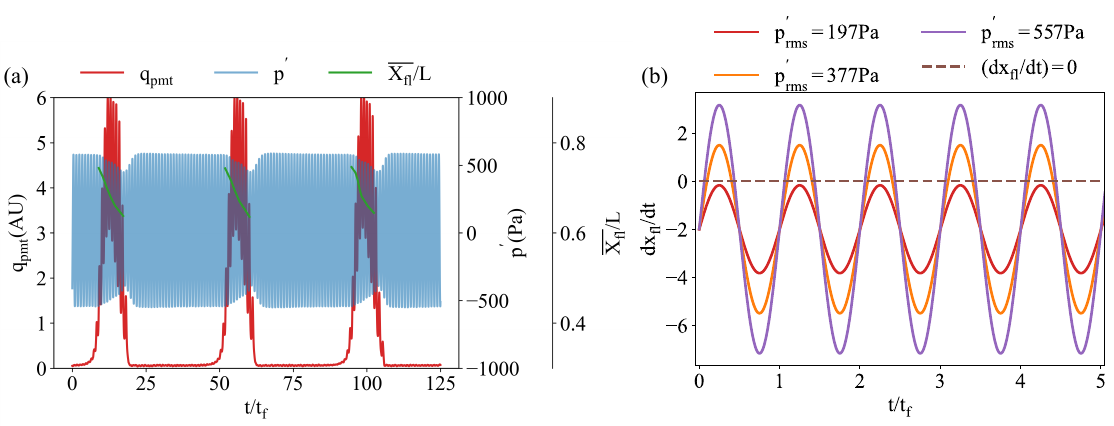}
    \caption{(a) Heat release rate ($q_{pmt}$) of a forced propagating flame at an rms pressure amplitude of 377 Pa, plotted alongside the measured pressure signal and the normalised mean flame position ($\overline{X_{fl}}/L$). Here, the flame has transitioned to the O-FREI regime. (b) Temporal variation of the model-predicted flame displacement rate ($dX_{fl}/dt$) for imposed rms pressure disturbances of 197 Pa, 377 Pa and 557 Pa.}
    \label{fig:PF_Limits}
\end{figure}

As the forcing amplitude is increased beyond $p^{\prime}_{rms} \geq 377$ Pa, the flame transitions to the oscillating FREI (O-FREI) regime (Fig. \ref{fig:PF_Limits}a).
It is interesting to examine this transition from a kinematic perspective. In the presence of external forcing (at the first harmonic), the mixture velocity immediately upstream of the flame front becomes $\bar{u} + u'(X_{fl},t)$, where $u'(X_{fl},t) = u_a(X_{fl}) \sin(\omega t)$. The flame displacement rate is therefore given by:
\begin{align}
\left( \frac{dX_{fl}}{dt} \right)_{f} = \bar{u} + u_a(X_{fl}) \sin(\omega t) - S_L = \left( \frac{dX_{fl}}{dt} \right)_{u} + u_a(X_{fl}) \sin(\omega t)
\tag{20}
\end{align}
The subscripts $u$ and $f$ in the above equation denote the flame displacement rate of the PF regime under unforced and forced conditions, respectively.
It is to be noted that ($\bar{u} - S_L$) in the above equation can be substituted with the measured displacement rate of the unforced propagating flame following Equation (18) and $u_a(X_{fl})$ can be evaluated using the formulation presented in Section.\ref{sec:vel_est}.
When forced at $p^{\prime}_{rms} = 197$ Pa, $\frac{dX_{fl}}{dt}$ remains negative at all times. Thus, the upstream propagation of the propagating flame is never arrested, and the system continues to exhibit the O-PF cycle, wherein the flame propagates upstream, extinguishes near $X=0$ (upstream constriction), and reignites again after a characteristic time period.

In contrast, for $p^{\prime}_{rms} = 377$ Pa, corresponding to the transitioned O-FREI regime, ${dX_{fl}}/{dt}$ becomes positive for a finite duration within each forcing cycle when the acoustic velocity is directed downstream (Fig. \ref{fig:PF_Limits}b). This indicates that the instantaneous convective transport of reactants toward the flame exceeds the rate at which the flame can propagate into the unburnt mixture. While this condition does not lead to immediate extinction, it represents a kinematic threshold beyond which the flame is subjected to repeated periods of adverse flow.
It is hypothesised that extinction arises through the cumulative effect of these periodic adverse flows.

Following ignition, near the heater, the flame resides in a region of elevated unburnt mixture temperature ($(X_{fl}/L) > 0.65$), resulting in a high local burning velocity and rapid chemical kinetics.
Here, the reduction in chemical timescale associated with elevated reactant temperature offsets the effect of the transient increase in convective transport during the downstream phase of the acoustic cycle (forcing directed along the X axis). Consequently, the flame remains robust and recovers during the subsequent upstream phase of the forcing cycle (forcing directed along the -X axis).
However, as the flame propagates upstream into regions of progressively lower wall temperatures ($(X_{fl}/L) < 0.65$), the unburnt mixture temperature decreases, leading to a reduction in the local burning velocity and a corresponding increase in the chemical timescale. 
As a result, the flame becomes increasingly susceptible to the imposed acoustic flow transients.
Beyond a critical upstream location, the residence time of reactants within the reaction zone during the downstream phase of the forcing cycle becomes insufficient to sustain complete combustion, leading to a temporary reduction in heat release rate. Although the flame partially recovers during the subsequent upstream phase of the forcing cycle, this recovery becomes progressively weaker as the flame propagates further upstream. Over successive acoustic cycles, this imbalance accumulates, and the flame eventually becomes unable to sustain the thermal structure required for combustion, resulting in extinction.
This mechanism explains why extinction does not occur immediately after ignition but instead occurs only after the flame has propagated into regions where the local thermochemical conditions render it sufficiently vulnerable to repeated flow-induced perturbations. Nevertheless, further studies are required to establish the detailed extinction mechanism.
Following extinction, the mixture auto-ignites after a time period that corresponds to the convective timescale of the reactant mixture to traverse from the extinction location to the ignition location. Effectively, the flame transitions into a regime exhibiting repetitive ignition-extinction dynamics alongside local oscillations at the forcing frequency, which corresponds to O-FREI. 

Consistent with the trends reported in Section \ref{subsec:SF}, following the transition to the O-FREI regime, increasing forcing pressure leads to a decrease in the mean heat release rate ($q_m$), while the amplitude of heat release rate fluctuations at the forcing time scale ($q_f$) tends to saturate (Fig. \ref{fig:PF_Trends}a). Likewise, the amplitude associated with the ignition–extinction instability ($q_{EI}$) and the corresponding Strouhal number ($St_{EI}$) decrease with increasing $p^{\prime}_{rms}$ (Fig. \ref{fig:PF_Trends}b). As discussed in Sections \ref{subsec:FR} and \ref{subsec:SF}, these trends can be interpreted in terms of asymmetric thermal loading, which explains the reductions in $q_m$, $q_{EI}$, and $St_{EI}$, and nonlinear saturation of flame front acceleration in the O-FREI regime at higher forcing pressures (following the dynamical model described in Equations 7 and 8), which accounts for the plateau in $q_f$.

\begin{figure}
    \centering
    \includegraphics[width=0.85\textwidth]{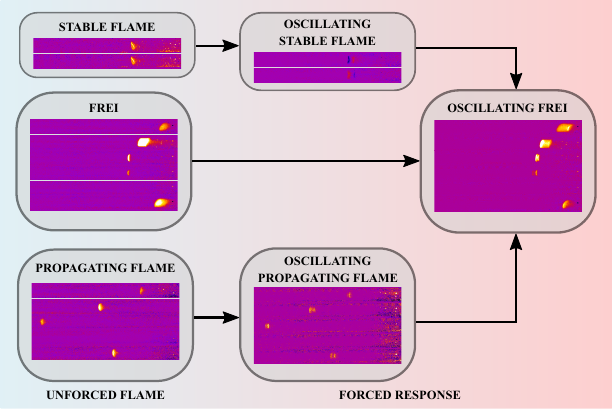}
    \caption{Schematic illustrating the transition route between different flame regimes under acoustic forcing.}
    \vspace{5mm}
    \label{fig:Transition_route}
\end{figure}

Fig. \ref{fig:Transition_route} summarises the dynamic response of different flame regimes to external acoustic forcing. Stable flames and propagating flames retain their qualitative behaviour while exhibiting local oscillations at the forcing frequency for an upstream forcing rms pressure amplitude of 197 Pa. However, as the forcing amplitude is increased to 377 Pa and above, the flames transition to the oscillating FREI (O-FREI) regime, in which the flame exhibits repeated cycles of ignition, propagation, extinction, and re-ignition while simultaneously undergoing local oscillations at the forcing frequency. In contrast, when an unforced FREI flame is subjected to acoustic forcing, it transitions directly to the O-FREI regime even at the lowest forcing rms pressure amplitude of 197 Pa.

\section{Conclusions} \label{sec:con}

This study examines the response of mesoscale flames to upstream acoustic forcing across three characteristic regimes — flames with repetitive extinction and ignition (FREI), stable flames (SF), and propagating flames (PF) — by systematically varying the forcing intensity at the first harmonic of the combustor tube. In the FREI regime, acoustic forcing drives the flame into an oscillating FREI (O-FREI) state, in which local oscillations at the forcing frequency are superimposed on the intrinsic FREI dynamics. For stable flames, forcing initially produces an oscillating stable flame (O-SF) response, characterised by back-and-forth flame-front motion and heat release oscillations about the mean value. With further increases in forcing amplitude, the flame transitions to O-FREI. Similarly, propagating flames subjected to low-amplitude forcing retain their baseline propagation behaviour while exhibiting enhanced thermoacoustic oscillations. This behaviour is termed the oscillating propagating flame (O-PF) regime. However, PF eventually transitions to O-FREI at higher forcing amplitudes.
Across all regimes, increasing forcing amplitude leads to a decrease in the mean heat release rate and a concurrent increase in the amplitude of heat release rate oscillations at the forcing frequency. In the O-FREI regime, to which all flames ultimately converge, both the peak heat release rate associated with the ignition–extinction cycle and the corresponding Strouhal number decrease with increasing forcing amplitude. These reductions are attributed to asymmetric thermal loading over a forcing cycle, whereas the amplification at the forcing frequency arises from direct velocity modulation at the flame front. The experimental trends are interpreted and supported using reduced-order mathematical models.

The identification of a universal O-FREI regime, to which all mesoscale flame regimes converge under sufficiently strong acoustic forcing, has important implications for mesoscale combustor design. In applications such as thermoelectric power generation, where spatially and temporally distributed wall heat loading is desirable, controlled access to the O-FREI regime offers a promising route for enhancing device performance and longevity.

\section*{Declaration of competing interest}
The authors report no conflict of interest.


\section*{Acknowledgments}
A. A. acknowledges support from the Prime Minister's Research Fellowship (PMRF) Scheme. S. B. acknowledges funding from the Indian National Academy of Engineering (INAE).

\appendix

\section{Mean Flow temperature of the reactant mixture}

The mean flow temperature, $T_m$, is estimated assuming steady heat transfer from the externally heated quartz tube to a fully developed laminar reactant stream. Axial conduction within the flow is neglected owing to a high Peclet number, while the wall temperature profile is assumed temporally invariant and radially uniform. Under these assumptions, the steady energy equation reduces to a balance between radial heat transfer and axial advection.

\begin{figure}[htbp]
    \centering
    \includegraphics[width=0.72\textwidth]{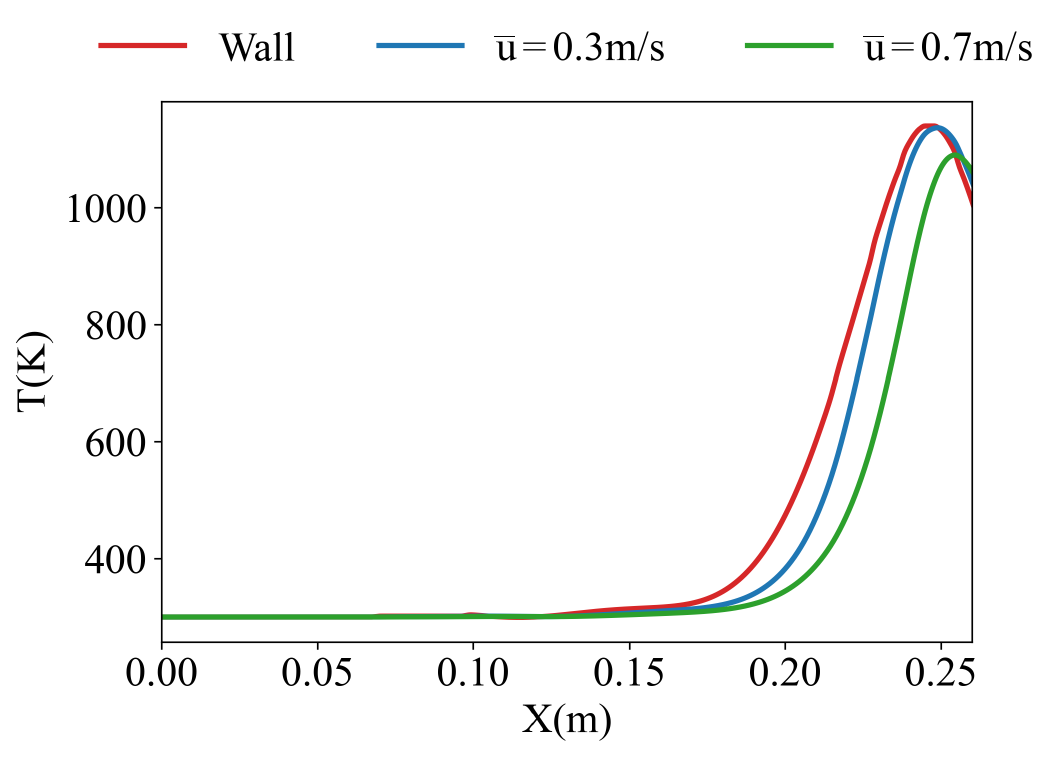}
    \caption{Mean flow temperature profile of the reactant mixture at mixture velocities of {0.3}{m/s} and {0.7}{m/s}. The velocity of {0.3}{m/s} corresponds to the FREI and PF regimes, whereas {0.7}{m/s} corresponds to the SF regime.}
    \label{fig:S1}
\end{figure}

Using a locally constant Nusselt number simplification ($Nu=3.66$) and performing an energy balance over an elemental control volume ($\Delta X$) yields,

\begin{equation}
q_w''(X)(2\pi r_i\Delta X)
=
\dot{m}\left[H(X+\Delta X)-H(X)\right],
\end{equation}

where $q_w''$ is the local wall heat flux, $r_i$ is the inner tube radius, $\dot{m}$ is the mass flow rate of the reactants, and $H$ is the specific enthalpy of the reactants. Simplifying the above equation yields the following spatial marching relation for the mean flow temperature,

\begin{equation}
T_m(X+\Delta X)
=
T_m(X)
+
\Delta X
\left(
\frac{\alpha Nu}{\bar{u}r_i^2}
\left[
T_o(X)-T_m(X)
\right]
\right).
\end{equation}

In the above equation, $\alpha$ is the thermal diffusivity of the mixture, and $T_o(X)$ is the measured wall temperature profile. The above equation is integrated axially from the known inlet temperature condition to obtain $T_m(X)$. For further details, the readers are referred to the work of Aravind et al. \cite{aravind_dynamics_2025}.

Figure \ref{fig:S1} presents the spatial profile of the mean flow temperature of the reactant mixture at mixture velocities of {0.3}{m/s} and {0.7}{m/s}, along with the corresponding wall temperature profile.

\bibliography{biblography}

@article{pizza_dynamics_2008,
	title = {Dynamics of premixed hydrogen/air flames in microchannels},
	volume = {152},
	number = {3},
	journal = {Combust. Flame.},
	author = {Pizza, Gianmarco and Frouzakis, Christos E. and Mantzaras, John and Tomboulides, Ananias G. and Boulouchos, Konstantinos},
	month = feb,
	year = {2008},
	pages = {433--450},
}

@article{maruta_micro_2011,
	title = {Micro and mesoscale combustion},
	volume = {33},
	number = {1},
	journal = {Proc. Combust. Inst.},
	author = {Maruta, Kaoru},
	month = jan,
	year = {2011},
	pages = {125--150},
	}

@article{fernandez-pello_micropower_2002,
	series = {Proceedings of the {Combustion} {Institute}},
	title = {Micropower generation using combustion: {Issues} and approaches},
	volume = {29},
	shorttitle = {Micropower generation using combustion},
	number = {1},
	journal = {Proc. Combust. Inst.},
	author = {Fernandez-Pello, A. Carlos},
	month = jan,
	year = {2002},
	pages = {883--899},
	}

@article{ju_microscale_2011,
	title = {Microscale combustion: {Technology} development and fundamental research},
	volume = {37},
	issn = {0360-1285},
	shorttitle = {Microscale combustion},
	number = {6},
	urldate = {2023-09-02},
	journal = {Prog. Energy Combust. Sci.},
	author = {Ju, Yiguang and Maruta, Kaoru},
	month = dec,
	year = {2011},
	pages = {669--715},
	}

@article{maruta_characteristics_2004,
	title = {Characteristics of {Microscale} {Combustion} in a {Narrow} {Heated} {Channel}},
	volume = {40},
	issn = {1573-8345},
	number = {5},
	urldate = {2023-09-04},
	journal = {Combust. Explos. Shock Waves.},
	author = {Maruta, K. and Parc, J. K. and Oh, K. C. and Fujimori, T. and Minaev, S. S. and Fursenko, R. V.},
	month = sep,
	year = {2004},
	pages = {516--523},
}

@article{fan_experimental_2009-1,
	title = {Experimental study of micro-scale premixed flame in quartz channels},
	volume = {32},
	number = {2},
	journal = {Proc. Combust. Inst.},
	author = {Fan, Yong and Suzuki, Yuji and Kasagi, Nobuhide},
	month = jan,
	year = {2009},
	pages = {3083--3090},
}

@article{di_stazio_combustion_2016,
	series = {Special issue on {Ninth} {Mediterranean} {Combustion} {Symposium}},
	title = {Combustion in micro-channels with a controlled temperature gradient},
	volume = {73},
	journal = {Exp. Therm Fluid Sci.},
	author = {Di Stazio, Annalisa and Chauveau, Christian and Dayma, Guillaume and Dagaut, Philippe},
	month = may,
	year = {2016},
	pages = {79--86},
	}

@article{aravind_dynamics_2025,
	title = {Dynamics of unsteady premixed flames in meso-scale channels and the effects of varying the wall heating conditions},
	volume = {1008},
	journal = {J. Fluid Mech.},
	author = {Aravind, Akhil and Vadlamudi, Gautham and Basu, Saptarshi},
	month = apr,
	year = {2025},
	pages = {A12},
}

@article{minaev_nonlinear_2007,
	title = {Nonlinear dynamics of flame in a narrow channel with a temperature gradient},
	volume = {11},
	number = {2},
	journal = {Combust. Theor. Model.},
	author = {Minaev, S. and Maruta, K. and Fursenko, R.},
	month = apr,
	year = {2007},
	pages = {187--203},
	}

@book{lieuwen_unsteady_2012,
	address = {Cambridge},
	title = {Unsteady {Combustor} {Physics}},
	publisher = {Cambridge University Press},
	author = {Lieuwen, Tim C.},
	year = {2012},
	}

@article{ZIMMERMANN2002285,
title = {Miniaturized flame ionization detector for gas chromatography},
journal = {Sens. Actuators, B},
volume = {83},
number = {1},
pages = {285-289},
year = {2002},
author = {Stefan Zimmermann and Peter Krippner and Albrecht Vogel and Jörg Müller},
}

@article{lloyd_burner_1974,
	title = {A burner for mixtures of very low heat content},
	volume = {251},
	number = {5470},
	journal = {Nature},
	author = {Lloyd, S. A. and Weinberg, F. J.},
	month = sep,
	year = {1974},
	pages = {47--49},
	}

@article{kang_versatile_2021,
	title = {A {Versatile} {Numerical} {Tool} for {Simulating} {Combustion} {Features} at {Small}-{Scales}},
	volume = {30},
	number = {1},
	journal = {J. Therm. Sci.},
	author = {Kang, Xin and Deng, Youcheng and Wang, Jianyong and Fan, Aiwu},
	month = jan,
	year = {2021},
	pages = {343--361},
	}

@article{maruta_characteristics_2005,
	title = {Characteristics of combustion in a narrow channel with a temperature gradient},
	volume = {30},
	number = {2},
	journal = {Proc. Combust. Inst.},
	author = {Maruta, Kaoru and Kataoka, Takuya and Kim, Nam Il and Minaev, Sergey and Fursenko, Roman},
	month = jan,
	year = {2005},
	pages = {2429--2436},
	}

@article{sheykhbaglou_numerical_2025,
	title = {Numerical {Analysis} of {Combustion} and {Thermal} {Performance} of a {Bluff}-{Body} and {Swirl}-{Stabilized} {Micro}-{Combustor} with {Premixed} {NH3}/{H2}/{Air} {Flames}},
	volume = {18},
	number = {4},
	journal = {Energies},
	author = {Sheykhbaglou, Soroush and Dimitriou, Pavlos and Sheykhbaglou, Soroush and Dimitriou, Pavlos},
	month = feb,
	year = {2025},
}

@article{wan_experimental_2012,
	series = {2011 {International} {Workshop} on {Molten} {Carbonates} \& {Related} {Topics}},
	title = {Experimental and numerical investigation on combustion characteristics of premixed hydrogen/air flame in a micro-combustor with a bluff body},
	volume = {37},
	number = {24},
	journal = {Int. J. Hydrogen Energy},
	author = {Wan, Jianlong and Fan, Aiwu and Maruta, Kaoru and Yao, Hong and Liu, Wei},
	month = dec,
	year = {2012},
	pages = {19190--19197},
	}

@article{noiray2008unified,
  title={A unified framework for nonlinear combustion instability analysis based on the flame describing function},
  author={Noiray, Nicolas and Durox, Daniel and Schuller, Thierry and Candel, S{\'e}bastien},
  journal={J. Fluid Mech.},
  volume={615},
  pages={139--167},
  year={2008},
  publisher={Cambridge University Press}
}

@article{ducruix2000theoretical,
  title={Theoretical and experimental determinations of the transfer function of a laminar premixed flame},
  author={Ducruix, S{\'e}bastien and Durox, Daniel and Candel, S{\'e}bastien},
  journal={Proc. Combust. Inst.},
  volume={28},
  number={1},
  pages={765--773},
  year={2000},
  publisher={Elsevier}
}

@article{resende_flame_2023,
	title = {Flame dynamics of hydrogen/air mixture in a wavy micro-channel},
	volume = {48},
	number = {36},
	journal = {Int. J. Hydrogen Energy},
	author = {Resende, P. R. and Ferrás, L. L. and Afonso, A. M.},
	month = apr,
	year = {2023},
	pages = {13682--13698},
	}

@article{gao_numerical_2023,
	title = {A numerical investigation of wavy microchannels with secondary branches under non-uniform heating},
	volume = {35},
	number = {2},
	journal = {Phys. Fluids.},
	author = {Gao, Zhigang and Zheng, Dawen and Bai, Junhua and Shang, Xiaolong and Li, Peng},
	month = feb,
	year = {2023},
	pages = {023610},
	}

@incollection{culick_combustion_1996,
	address = {Dordrecht},
	title = {Combustion {Instabilities} in {Propulsion} {Systems}},
	booktitle = {Unsteady {Combustion}},
	publisher = {Springer Netherlands},
	author = {Culick, F. E. C.},
	year = {1996},
	pages = {173--241},
}

@article{aravind_response_2025,
	title = {Response of non-premixed jet flames to blast waves},
	volume = {1013},
	journal = {J. Fluid Mech.},
	author = {Aravind, Akhil and Vadlamudi, Gautham and Basu, Saptarshi},
	month = jun,
	year = {2025},
	pages = {A3},
}

@article{mohan_self-excited_2023,
	title = {Self-excited intermittent thermoacoustic fluctuations in an annular combustor exhibiting flame transient phenomena: {Physical} mechanisms and modeling},
	volume = {35},
	journal = {Phys. Fluids.},
	author = {Mohan, Balasundaram and Mariappan, Sathesh},
	month = nov,
	year = {2023},
	pages = {114108},
}

@article{Flores-Montoya_Muntean_Sánchez-Sanz_Martínez-Ruiz_2022, 
    title={Non-adiabatic modulation of premixed-flame thermoacoustic frequencies in slender tubes}, 
    volume={933}, 
    journal={J. Fluid Mech.}, 
    author={Flores-Montoya, Enrique and Muntean, Victor and Sánchez-Sanz, Mario and Martínez-Ruiz, Daniel}, 
    year={2022}, 
    pages={A50}
}

@book{munjal1987acoustics,
  title={Acoustics of ducts and mufflers with application to exhaust and ventilation system design},
  author={Munjal, Manchar Lal},
  year={1987},
  publisher={John Wiley \& Sons}
}

@book{parker2012practical,
  title={Practical numerical algorithms for chaotic systems},
  author={Parker, Thomas S and Chua, Leon},
  year={2012},
  publisher={Springer Science \& Business Media}
}

\end{document}